\documentclass[offcenter,tocnosub,noragright,centerchapter,12pt]{uiucecethesis09}

\usepackage[utf8]{inputenc}
\usepackage[english]{babel}

\addto\captionsenglish{
	\renewcommand{\bibname}{References}
}

\makeatletter

\usepackage{setspace}
\usepackage{epsfig}  
\usepackage{amsmath}  
\usepackage{lscape}  
\usepackage[justification=raggedright]{caption}	

\usepackage{fullpage} 
\usepackage{mathtools}
\usepackage{amssymb}  
\usepackage{color}    
\usepackage{float}    
\usepackage{algorithm}
\usepackage{algpseudocode}
\usepackage{framed}   
\usepackage{enumerate}  
\usepackage{dsfont}  
\usepackage{graphicx}
\usepackage{subcaption}
\usepackage{url}

\DeclarePairedDelimiter\angles{\langle}{\rangle}

\DeclarePairedDelimiter\floor{\lfloor}{\rfloor}

\newcommand{\cD}{\mathcal{D}}

\newcommand{\cS}{\mathcal{S}}

\newcommand{\sR}{\mathbb{R}}
\newcommand{\E}{\mathbb{E}}

\newcommand{\dd}{\mathrm{d}}

\newcommand{\Var}{\mathrm{Var}}

\newcommand{\Geo}{\mathrm{Geo}}

\newcommand{\ML}{\mathrm{ML}}

\newcommand{\Poisson}{\mathrm{Poisson}}
\newcommand{\Binom}{\mathrm{Binom}}
\newcommand{\LL}{\mathrm{LL}}
\newcommand{\vLL}{\mathrm{vLL}}
\newcommand{\HLL}{\mathrm{HLL}}
\newcommand{\Zipf}{\mathrm{Zipf}}
\newcommand{\WSE}{\mathrm{WSE}}

\newcommand{\lb}{\mathrm{lb}}
\newcommand{\ub}{\mathrm{ub}}
\newcommand{\middleone}{\mathrm{mid}_1}
\newcommand{\middletwo}{\mathrm{mid}_2}

\newcommand{\ind}{\mathds{1}}

\msthesis

\title{Per-Flow Cardinality Estimation Based On Virtual LogLog Sketching}
\author{Zeyu Zhou}
\department{Electrical and Computer Engineering}
\degreeyear{2016}

\advisor{Professor Bruce Hajek}

\begin{document}

%

%
\maketitle

\parindent 1em%

\frontmatter

%
\begin{abstract}
Flow cardinality estimation is the problem of estimating the number of distinct elements in a data flow, often with a stringent memory constraint. It has wide applications in network traffic measurement and in database systems. The virtual LogLog algorithm proposed recently by Xiao, Chen, Chen and Ling estimates the cardinalities of a large number of flows with a compact memory. The purpose of this thesis is to explore two new perspectives on the estimation process of this algorithm. Firstly, we propose and investigate a family of estimators that generalizes the original vHLL estimator and evaluate the performance of the vHLL estimator compared to other estimators in this family. Secondly, we propose an alternative solution to the estimation problem by deriving a maximum-likelihood estimator. Empirical evidence from both perspectives suggests the near-optimality of the vHLL estimator for per-flow estimation, analogous to the near-optimality of the HLL estimator for single-flow estimation.

\end{abstract}

%

%
\begin{acknowledgments}
I thank Prof. Bruce Hajek, for introducing this topic to me and guiding me through the completion of this thesis with great patience. I thank my parents, for their unconditional love and support.
\end{acknowledgments}

%
\tableofcontents

%

%

%


%

\mainmatter

%

\chapter{Introduction} \label{ch:introduction}

\section{Flow cardinality estimation}

Today's Internet is flooded with data. The measuring and monitoring of Internet data traffic has led to many useful applications but it is also challenging due to the sheer volume and speed of the traffic. \emph{Flow cardinality estimation} is one of the fundamental problems in network traffic measurement and it is the problem addressed by this thesis. 

In general, we define \emph{cardinality estimation} to be the problem of estimating the number of \emph{distinct} elements $n$ of a given multiset\footnote{A multiset is a generalization of a set, with the difference that a multiset can have duplicate elements.} $\cS$. For example, if $\cS = \{1,2,3,2,1\}$, then $n = 3$. The problem is often considered in the stream model --- we are only allowed to observe each element in $\cS$ once, then the element is discarded forever. 

Flow cardinality estimation, in its essence, is cardinality estimation in the context of network traffic measurement, where we define a \emph{flow} to be a multiset of data packets defined by certain properties observed on a network link over a period of time. The purpose of flow cardinality estimation is to estimate the number of \emph{distinct} packets in the multiset, where the distinction is made based on some properties such as values of certain packet header fields of interest. 

For example,  we can consider a \emph{per-source flow} that contains all the packets from the same \emph{source address} and use cardinality estimation to estimate the number of \emph{distinct destination addresses} among them. In this case, the particular source address is known as the flow's identifier (\emph{flow ID}). Some commonly used flow identifiers are source/destination address, source/destination port number, protocol type, or a combination of them (such as source-destination pair).

\bigskip
\textbf{Challenges:} The challenges of flow cardinality estimation are mainly twofold: time and space constraints. First of all, high data transfer rates in some communication networks today make it infeasible to spend much time on processing each element. According to \cite{Giroire2009} and \cite{ChassaingGerin2011}, on a typical OC-768\footnote{OC is an acronym for \emph{Optical Carrier}. OC-n specifies the transmission rate of digital signals on a Synchronous Optical Networking (SONET) fiber optic network, equivalent to $n \times 51.84$ Mbit/s.} backbone network link with 40 Gbps traffic speed, the time available to process each packet is at best about tens of nanoseconds, corresponding to no more than a hundred elementary operations. 

Secondly, such a high traffic rate necessitates the need to use on-chip cache memory on network processors, in order to achieve real-time processing and maintain high throughput. But most on-chip caches on processors are made of SRAM, typically only a few megabytes \cite{Xiaoetc2015}. A na\"{i}ve approach that maintains a look-up dictionary to record all the distinct elements seen thus far is too costly, since it requires $O(n)$ bits. In some applications, $n$ can be on the order of billions.

Therefore, an ideal algorithm should be simple and quick in processing each flow element and be compact in memory usage. The good news is that it is usually not necessary to know the exact value of a flow's cardinality. Many approximation algorithms have been developed to explore the trade-off between estimation accuracy and space/timing efficiency. 

\bigskip
\textbf{Applications Examples:} One important direction of application of cardinality estimation is \emph{network anomaly detection}. Estan et al. \cite{EstanVargheseFisk2003} suggested three such applications: detecting port scans, detecting denial of service (DoS) attacks and estimating worm spread rate. 

Let us take port scan detection as an example to illustrate the idea. A port scan is a probing process that sends service requests from a client to a range of port addresses of a server with the aim of finding an open port, which an attacker can take advantage of to find vulnerabilities on the server. A router can detect such a process by keeping track of packet flows and using cardinality estimation to measure the number of distinct port addresses attempted by each source. Any source that is trying to connect to an abnormally large number of port addresses in a short time interval should be suspected of conducting a port scan. 

Other than network anomaly detection, we can apply flow cardinality estimation to many other problems that have a \emph{count-distinct} nature by adapting the concept of a flow to contain other (maybe abstract) types of data. For example, Google might be interested in estimating the number of distinct users that query certain keywords in a period of time \cite{Xiaoetc2015}. Here a \emph{per-keyword} flow can be defined to be all the user IP addresses that query that keyword. Information about the cardinalities of such flows reflects the popularity of those keywords which might be helpful in the optimization of Google's database and search algorithm. Much research on cardinality estimation was motivated by database-related applications (e.g. \cite{FlajoletMarin1985} and \cite{HeuleNunkesserHall2013}).

\section{Motivation and contributions}

State-of-the-art algorithms for cardinality estimation comprise two processes. First, the \emph{sketching process} reads elements from the flow and stores useful information in a compact data structure. Second, in the \emph{estimation process}, an estimator takes the recorded information as input and outputs the estimated cardinality of the flow. 

The \emph{HyperLogLog} (HLL) algorithm proposed by Flajolet et al. in \cite{Flajoletetc2007} is near-optimal and has been widely adopted in many industry systems due to its simplicity and excellent performance; according to \cite{Xiaoetc2015}, it is the best existing algorithm for single-flow estimation. Roughly speaking, the HLL algorithm requires hundreds of bytes to make a fair estimation for a single flow with cardinality up to $4 \times 10^9$. Details of the HLL algorithm are presented in Section \ref{subsec:ll_hll}. 

However, in some applications the cardinalities of multiple flows need to be estimated at the same time --- this is known as per-flow cardinality estimation. In this case, the rate of hundreds of bytes per flow is still too much in some important scenarios where the number of flows can be on the order of tens of millions and memory usage is critical. To this end, Xiao et al. \cite{Xiaoetc2015} proposed the \emph{virtual LogLog} algorithm based on memory sharing at register level, which can potentially bring down the memory cost from \emph{hundreds of bytes} per flow to \emph{one bit} per flow on average. Details of the sketching process of this algorithm are described in Section \ref{subsec:vLL_sketching}. The estimation process of this algorithm uses an estimator called vHLL, summarized in Section \ref{subsec:review_vHLL}.

The main purpose of this thesis is to propose and investigate alternative estimators for the estimation process of the virtual LogLog algorithm in \cite{Xiaoetc2015}:

\begin{itemize}
	\item The original vHLL estimator of \cite{Xiaoetc2015} is based on the HLL estimator for single-flow estimation. We show that the HLL estimator can be generalized by a family of estimators parametrized by a value $\theta$ (Section \ref{sec:ll_theta}); for HLL, $\theta = -1$. The idea of this generalization can be easily applied to the vHLL estimator for per-flow estimation as well. Although it is already known that for single-flow estimation, $\theta = -1$ (i.e. HLL) is near-optimal \cite{Flajoletetc2007}, it is not clear whether this near-optimality at $\theta = -1$ extends to the case of per-flow estimation. We provide empirical evidence to show that indeed $\theta = -1$ (i.e. vHLL) is still optimal for per-flow estimation. 
	\item We introduce an alternative approach to the estimation problem -- a maximum-likelihood estimator. We find that this estimator does not significantly improve the estimation process, compared to the vHLL estimator.
\end{itemize}

Empirical evidence from both perspectives suggests the near-optimality of the vHLL estimator for per-flow estimation, on the basis of the virtual LogLog algorithm \cite{Xiaoetc2015}.

\section{Thesis overview}

The rest of this thesis is organized as follows. Chapter \ref{ch:preliminaries} provides the background knowledge and preliminaries to the estimators discussed in this thesis. Chapter \ref{ch:generalize_theta} and Chapter \ref{ch:mle} investigate alternative estimators of the virtual LogLog algorithm from the aforementioned two perspectives. Chapter \ref{ch:generalize_theta} introduces the $\vLL_\theta$ estimator, a generalization of the vHLL estimator, and evaluates its performances for different values of $\theta$. Chapter \ref{ch:mle} derives the maximum-likelihood estimator and compares its performance with that of the $\vLL_\theta$ estimator. Chapter \ref{ch:conclusion} summarizes the conclusions of the thesis and points to possible future works.

\chapter{Background and Preliminaries} \label{ch:preliminaries}

This chapter briefly overviews cardinality estimation algorithms and points to relevant references. The focus is on summarizing previous works that are essential to the understanding of the main body of this thesis: the LogLog \cite{DurandFlajolet2003} and HyperLogLog \cite{Flajoletetc2007} algorithms for single-flow estimation and the virtual LogLog algorithm \cite{Xiaoetc2015} for per-flow estimation. 

\section{Single-flow cardinality estimation} \label{sec:single_flow_estimation}

Given a single flow of elements $x_1, x_2, x_3, \cdots$, the goal of cardinality estimation is to estimate the number of distinct elements in the flow. There are two general approaches to this problem: sampling and streaming. A sampling-based algorithm collects a small sample of elements from the flow while bypassing other elements and infers the cardinality of the flow from sample information. A streaming-based algorithm, in contrast, processes every element in the flow. Most state-of-the-art algorithms are based on streaming. For a review of these two kinds of algorithms and a comparison between them, see \cite{Gibbons2007}.

A streaming-based algorithm for cardinality estimation has three integral components:
\begin{itemize}
    \item A \emph{hash function} $H$, mapping each element $x_i$ of the flow into a hashed value $H(x_i)$ of a chosen type. The purpose of $H$ is generally twofold: filtering out duplicate elements (relying on the fact that multiple appearances of the same element have the same hash) and providing randomization. 
    \item A \emph{compact data structure} $\cD$, capturing certain low-dimensional statistical information of the flow's cardinality by recording features of the hashed values. In some literature, the data structure is known as a \emph{sketch}, meaning it is a concise summary of the flow. The process of updating the sketches is called \emph{sketching}.
    \item An \emph{estimator} $E$ that takes the content of $\cD$ as input and outputs the estimated cardinality, thus $\widehat{n} = E(\cD)$. 
\end{itemize}

Usually we want the estimator to be unbiased (i.e. $\E[\widehat{n}|n] = n$). One commonly used metric for the evaluation of the performance of an estimator is the \emph{relative standard error} $\triangleq \frac{\sqrt{\Var(\widehat{n})}}{n}$. For two unbiased estimators using the same amount of memory, the one with a smaller relative standard error is better. For a detailed classification and comparison of existing single-flow cardinality estimation algorithms, see \cite{MetwallyAgrawalAbbadi2008, CliffordCosma2010}. We take a closer look at the LogLog and HyperLogLog algorithms in Section \ref{subsec:ll_hll}. 

\subsection{LogLog and HyperLogLog} \label{subsec:ll_hll}

The HyperLogLog (HLL) \cite{Flajoletetc2007} algorithm is the best-known algorithm in practice, due to its simplicity and good performance. HLL is a successor of LogLog \cite{DurandFlajolet2003}: the two algorithms use the same hash function and data structure (therefore have the same sketching process) and only differ in their choice of estimators. 

The data structure of the LogLog/HLL algorithm is a single array of $k$ registers (counters), denoted as $S$, where $S[i]$ refers to the $i^{th}$ register in the array (or the value stored in it, depending on the context). Suppose we have a hash function $H$ that maps each element of the flow to a sufficiently long bit string. Define a function $\rho$ that takes a bit string and outputs the position of the leftmost $1$-bit of the string, e.g. $\rho(000010\cdots) = 5$. The sketching process of LogLog/HLL is summarized as follows.
\begin{framed}
\textbf{LogLog/HLL sketching process}:
\begin{enumerate}
    \item Initialize all the registers in $S$ with value $0$. Let $l = \log_2 k$. 
    \item For each element $x$ in flow:
    \begin{enumerate}[(i)]
        \item $\angles{b_1 b_2 b_3 \cdots} \gets H(x)$. Hash $x$ into a sufficiently long bit string. 
        \item Divide the bit string into two parts:
            \begin{itemize}
                \item $j \gets \angles{b_1 b_2 \cdots b_l}$. Use the prefix $l$ bits to select a register in $S$. \item $q \gets \angles{b_{l+1} b_{l+2} \cdots}$. Use the remaining bits to update the selected register. 
            \end{itemize}
        \item $S[j] \gets \max\{S[j], \rho(q)\}$. Update the value of $S[j]$ by the larger of the current value of $S[j]$ and the position of the leftmost $1$-bit of $q$.
    \end{enumerate}
\end{enumerate}
\end{framed}
Here are a few points to note about the above sketching process. First, $k$ is chosen to be some integer power of $2$ so that $l$ is an integer and $j \in \{0,1,\cdots,k-1\}$. Assume that for a random element $x$, $H(x)$ is a random bit string with independent and uniform bits. Then $j$ can be regarded as a uniform random variable in $\{0,\cdots,k-1\}$, which means that the $k$ registers in $S$ are equally likely to be selected for updating for a randomly given element.

Second, the values of the registers in $S$ after the sketching process should not be affected by duplicate elements, nor by the order of the elements appearing in the flow. This is guaranteed by hashing and the max operation in step (iii). 

Third, how large can a register's value get? With the previous assumption on the independence and uniformity of the bits in the hash output string, for a random element $x$ we have $\rho(q) = i$ with probability $\frac{1}{2^i} (i \geq 1)$. That is, $\rho(q)$ can be regarded as a geometric random variable with parameter $1/2$. By the same assumption, we expect about $\frac{n}{k}$ distinct elements distributed to any particular register $S[j]$. Suppose that this number $\frac{n}{k}$ is exact, then at the end of the sketching process, $S[j]$ is just the maximum of $\frac{n}{k}$ independent $\Geo(1/2)$ random variables. According to \cite{DurandFlajolet2003}, previous study has shown that the expectation of $S[j]$ is close to $\log \frac{n}{k}$ with a small additive bias. Therefore, on average each register needs about $\log \log \frac{n}{k} = O(\log \log n)$ bits to store its value --- this is why such a register is also known as a \emph{LogLog sketch}. Based on this discussion, a rough estimator for $n$ can be $\widehat{n} = k 2^{S[j]}$. Following this idea, the LogLog estimator \cite{DurandFlajolet2003} replaces $S[j]$ with the average of the $k$ register values: 
\begin{equation} \label{eq:loglog_estimator}
    \widehat{n} = \LL(S) \triangleq \eta k 2^{\frac{S[0] + \cdots + S[k-1]}{k}},
\end{equation}
where $\eta$ is a suitable bias correction factor. The HLL algorithm \cite{Flajoletetc2007} uses a different estimator:
\begin{equation} \label{eq:hll_estimator}
    \widehat{n} = \HLL(S) \triangleq \gamma k \frac{k}{2^{-S[0]} + \cdots + 2^{-S[k-1]}}
\end{equation}
with a different bias correction factor $\gamma$. Note that $2^{\frac{S[0] + \cdots + S[k-1]}{k}}$ is commonly known as the \emph{geometric mean} of the values $2^{S[0]}, \cdots, 2^{S[k-1]}$, while $\frac{k}{2^{-S[0]} + \cdots + 2^{-S[k-1]}}$ is the \emph{harmonic mean} of those values. 

We will introduce two other estimators for single-flow estimation in Section \ref{sec:ll_theta} ($\LL_\theta$ estimator) and Section \ref{sec:mle_single_flow} (maximum-likelihood estimator), respectively.

It has been shown in \cite{DurandFlajolet2003} and \cite{Flajoletetc2007} that the LogLog and HLL estimators are \emph{asymptotically approximately unbiased} in the sense that, as $n \rightarrow \infty$, $\frac{\E[\widehat{n}]}{n}$ is very close to $1$ with a practically negligible fluctuation. It has been shown that as $n \rightarrow \infty$, the bias correction constants are independent of $n$. They do depend on $k$, however; but as $k$ gets large (e.g. $k \geq 64$), they can be safely replaced by constants for all practical purposes. Practical values of $\eta$ and $\gamma$ are $0.39701$ and $0.7213$ respectively.

The \emph{relative standard error} $\triangleq \frac{\sqrt{\Var(\widehat{n})}}{n}$ is approximately $\frac{1.30}{\sqrt{k}}$ for the LogLog estimator \cite{DurandFlajolet2003} and approximately $\frac{1.04}{\sqrt{k}}$ for the HLL estimator \cite{Flajoletetc2007}, as $n \rightarrow \infty$. Therefore, using the same number of registers (same amount of memory), HLL's estimate is more accurate than that of LogLog. With the HLL algorithm, we can achieve an estimate accuracy (in terms of relative standard error) of about $5\%$ by choosing $k = 512$. If we use $5$ bits for each register (for measuring cardinalities up to $2^{2^5} \approx 4 \times 10^9$), then in total we need $5k = 2560$ bits $= 320$ bytes for one estimation. 

According to Section 4 of \cite{Flajoletetc2007}, the HLL algorithm is near-optimal, in the sense that its relative standard error $\frac{1.04}{\sqrt{k}}$ is quite close to the lower bound $\frac{1}{\sqrt{k}}$ for a wider class of algorithms based on order statistics. For a more detailed discussion of this lower bound, see \cite{ChassaingGerin2011}.

One problem of the HLL algorithm is that it is highly biased and inaccurate for the estimation of small cardinalities. Methods to remedy this flaw are considered in Section 4 of \cite{Flajoletetc2007} and further in \cite{HeuleNunkesserHall2013}.

\section{Per-flow cardinality estimation with memory sharing}

In many real-world applications we need to estimate the cardinalities of multiple flows at the same time. For example, consider a stream of data packets from many flows observed by a network monitor device (such as a router). Let each packet be abstracted as a 2-tuple $(f,x)$, where $f$ is the IP source address of the packet (the flow ID) and $x$ is the destination address (i.e. an element of the flow). The goal is, at the end of the measuring period, to estimate the number of distinct destination addresses from each given source address, i.e. the cardinality of each flow. For example, if the stream of packets is $(A,2), (C,9), (A,3), (A,2), (C,1), (B,8)$, then flow $A$ is $\{2,3,2\}$ with cardinality $2$, flow $B$ is $\{8\}$ with cardinality $1$ and flow $C$ is $\{9,1\}$ with cardinality $2$.

\begin{figure}[!h]
	\centering
	\includegraphics[scale=0.5]{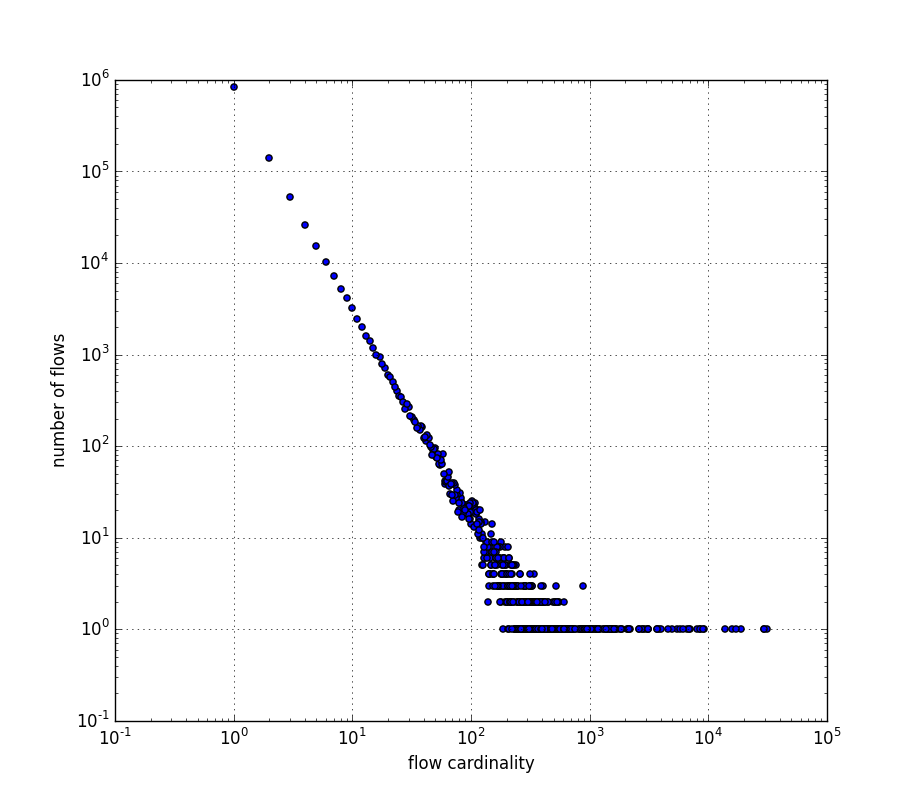}
	\caption{A typical flow cardinality distribution from real trace files. x-axis: flow cardinality. y-axis: the number of flows with the corresponding cardinality. Total number of flows: $1,116,535$. Average flow cardinality: $2.52$.}
	\label{fig:trace_flow_cardi_dist}
\end{figure}

An immediate idea to solve this problem is to allocate a separate block of memory for each flow and use any of the existing single-flow cardinality estimation algorithms for each flow. Since we do not know the cardinalities of the flows beforehand, it is inevitable to allocate the maximum amount of memory for each flow; with the HLL algorithm, we still need hundreds of bytes per flow. However, in many applications most of the flows have small cardinalities. Figure \ref{fig:trace_flow_cardi_dist} shows an example of flow cardinality distribution from real-world data.\footnote{Data are retrieved from network traffic trace files \cite{CAIDA} recorded on a backbone link between San Jose and Los Angeles during a ten minute interval.} Here, a flow is defined by an IP source address. The cardinality of a flow is the number of distinct destination addresses among all the packets in the flow. In this example, the majority (about $90\%$) of the flows have cardinality of only $1$, while only six flows have cardinalities larger than $10^4$.

In view of this waste of memory, some algorithms have been proposed that allow memory to be shared among flows. Such an algorithm usually combines the following components/ideas:
  
\begin{itemize}
	\item \emph{A memory pool}, which is the actual source of memory for all estimations.
	\item \emph{A virtual data structure for each flow}. This data structure does not physically exist but is logically constructed by (often randomly) pointing to memory units in the pool. Consequently, difference flows may share some common parts of the memory. 
	\item  In the sketching process, for each incoming element, the algorithm applies the sketching process of an existing single-flow estimation algorithm to update the virtual data structure allocated for the corresponding flow. 
	\item After the sketching process, the algorithm uses an estimator to estimate the cardinality of any given flow (with its ID). The estimator needs to take into account the interference among flows due to memory sharing. This estimation process can be done off-line (in comparison to the online sketching process).
\end{itemize}

Such algorithms may differ from each other in any of the above aspects. A review and comparison of some per-flow cardinality estimation algorithms based on memory sharing can be found in \cite{Xiaoetc2015}. All these algorithms, however, share memory at \emph{bit} level (meaning that the basic unit of the memory pool and of the virtual data structure is a bit), which \cite{Xiaoetc2015} claims to be inherently too noisy. 

Another algorithm that shares memory at bit level, which \cite{Xiaoetc2015} did not mention, is the \emph{virtual FM sketches} algorithm proposed in \cite{Moetal2014}. The algorithm constructs multiple virtual bit arrays from a bit pool for each flow and relies on the sketching process of the FM sketches\footnote{See \cite{FlajoletMarin1985} by Flajolet and Marin for a description of the FM sketches algorithm for single-flow estimation.} algorithm for single-flow estimation. The algorithm adopts a maximum-likelihood estimator in the estimation process based on bit patterns in the sketches. We will also consider a maximum-likelihood estimator solution in Chapter \ref{ch:mle}, but for a different sketching scheme. The performance reported in \cite{Moetal2014} shows that with memory cost of 1 bit per flow, the algorithm can estimate flows with cardinalities up to $3000$ and average flow cardinality about $2.5$; it can achieve relative standard error of about $20\%$ for flows with cardinalities less than $500$ and $10\%$ for flows with cardinalities in the range $500$ to $3000$. For relatively small flows with cardinalities in the range $100$ to $500$, this algorithm may outperform the virtual LogLog algorithm with the vHLL estimator \cite{Xiaoetc2015}; but it does not appear to be as competitive for larger ranges of flow cardinality. 

In view of the inherent drawback of bit-level sharing, \cite{Xiaoetc2015} proposed the \emph{virtual LogLog} algorithm based on memory sharing at \emph{register} level and showed through experiments that such an algorithm outperforms previous ones. The data structure and sketching process of this algorithm are presented in the following subsection. 

\subsection{Virtual LogLog register sharing and sketching process} \label{subsec:vLL_sketching}

\begin{figure}[H]
    \centering
    \includegraphics[scale=0.5]{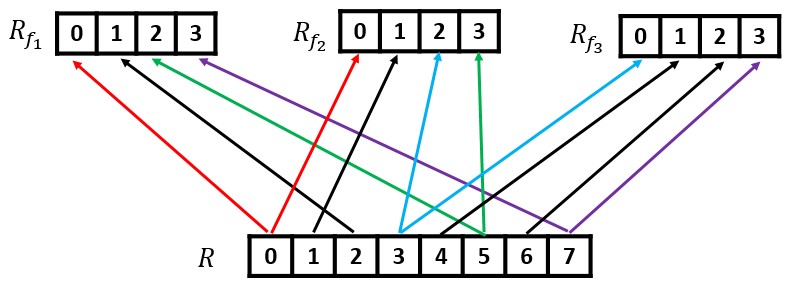}
    \caption{An example of the register pool and virtual register arrays. Here $m = 8$, $k = 4$. There are three flows and three corresponding virtual register arrays: $R_{f_1} = \left[R[0], R[2], R[5], R[7]\right]$, $R_{f_2} = \left[R[0], R[1], R[3], R[5]\right]$, $R_{f_3} = \left[R[3], R[4], R[6], R[7]\right]$.}
    \label{fig:virtual_estimator_example}
\end{figure}

\textbf{Virtual Data Structure:} The virtual LogLog algorithm \cite{Xiaoetc2015} keeps a memory pool in the form of a register array, denote as $R$. Denote the size of the array by $m$, which is usually a very large number. $R[j]$ refers to the $j^{th}$ register in the array (or the value stored in the register, depending on the context). For each flow $f$, we form a virtual data structure (\emph{virtual register array}) denoted as $R_f$, which is a logically constructed array of $k$ registers with the $i^{th}$ register denoted as $R_f[i]$. The registers of $R_f$ are randomly selected from $R$ by using $k$ independent hash functions $G_0, G_1, \cdots, G_{k-1}$, each mapping the flow ID uniformly to an integer in $\{0, \cdots, m-1\}$, i.e. 
\begin{equation} \label{eq:register_mapping_from_pool}
    R_f[i] = R[G_i(f)], \quad 0 \leq i \leq k-1.
\end{equation}
The $k$ hash functions can be implemented by a single master hash function $G$ as follows:
\begin{equation} \label{eq:master_hash}
    G_i(f) = G(f | i),
\end{equation}
where ``$|$" is the concatenation operator. It should be emphasized that $R_f$ does not need to be physically constructed (thus it is ``virtual"). A simple example is shown in Figure \ref{fig:virtual_estimator_example} to illustrate the concepts. In the example, say we want to update the third register of flow $f_2$, i.e. $R_{f_2}[3]$, what actually will be updated is $R[5]$ --- we do not even need to know where the other registers in $R_{f_2}$ are. 

A caveat should be pointed out here, which the original paper \cite{Xiaoetc2015} omitted: it could happen that two logically distinct registers of a virtual register array are mapped from the same physical register in the pool, i.e. $G_i(f) = G_j(f)$ for some $i \not= j$. Certainly we wish to avoid this situation, but it is tolerable if the number of such ``collisions" is very small. 

To see how likely it is for such a ``collision" to occur, we can consider a bins-and-balls analogous problem: suppose we have $k$ balls and $n$ bins and we throw the balls sequentially, independently and uniformly at random into the bins. In the end how many bins will contain more than one balls? If $k$ is reasonably large, then the number of balls distributed at \emph{each} bin can be approximated by an \emph{independent} Poisson random variable $X$ with mean $\frac{k}{m}$ (see Chapter 5.4 of \cite{MitzenmacherUpfal}). So we have
\begin{equation*}
    \Pr\{X \geq 2\} = 1 - \Pr\{X \leq 1\} \approx 1 - \left(1+\frac{k}{m}\right)e^{-\frac{k}{m}} \approx 1 - \left(1+\frac{k}{m}\right)\left(1-\frac{k}{m}\right) = \frac{k^2}{m^2},
\end{equation*}
where the last approximation assumes $k \ll m$. Therefore out of the $m$ bins, we expect to have approximately $\frac{k^2}{m}$ bins that hold more than one ball. Ideally we want $\frac{k^2}{m}$ to be small, which means with high probability the registers in a virtual register array are all mapped from distinct physical registers in the pool. This factor should be included in the design consideration.  

\textbf{Sketching Process:} The sketching process of virtual LogLog is \emph{almost} identical to that of LogLog/HLL (recall from Section \ref{subsec:ll_hll}): for each packet $(f,x)$ in the stream, we process $x$ to obtain $j$ (the register selector) and $\rho(q)$ (to be compared with the selected register's value); except that the last step (iii) becomes
\begin{equation*}
    R_f[j] \leftarrow \max\{R_f[j], \rho(q)\}.
\end{equation*}
That is, we are treating the virtual register array as the actual register array. By combining (\ref{eq:register_mapping_from_pool}) and (\ref{eq:master_hash}), the above expression can be re-written as 
\begin{equation} \label{eq:vLL_sketching_max_update}
    R[G(f|j)] \gets \max\{R[G(f|j)], \rho(q)\}.
\end{equation}
Again, it shows that updates are actually made in the physical registers in $R$. It should now be clear why the algorithm is called virtual LogLog: it is based on virtual register arrays, where each register is a LogLog sketch. 

After the sketching process, we obtain $m$ register values. The remaining problem is to infer flow cardinalities from these register values: this is the \emph{estimation process}. We will consider two kinds of estimators in Chapters \ref{ch:generalize_theta} and \ref{ch:mle}, respectively.

\section{A per-flow estimator performance metric -- Weighted square error} \label{sec:wse}

Before we discuss specific estimators for per-flow cardinality estimation, we consider a metric based on weighted square error to evaluate the performance of any given estimator. 

Suppose that the incoming data stream contains $M$ flows, with cardinalities $n_1, n_2, \cdots, n_M$. Given any estimator $E$, suppose its corresponding estimates for the flows' cardinalities are $\widehat{n}_1, \widehat{n}_2, \cdots, \widehat{n}_M$. The \emph{weighted square error} of estimator $E$ is defined as
\begin{equation}
    \WSE(E) \triangleq \sum_{i=1}^M \left(n_i - \widehat{n}_i\right)^2 w(n_i),
\end{equation}
where $w(\cdot)$ is a weight function mapping the cardinality of a flow to a positive number. For two estimators using the same amount of memory, the one with a lower WSE is better. 

\subsection{Weight function}

The choice of the weight function depends on the specific application. For example, if each flow is considered equally important regardless of its cardinality, then we can simply let $w(n) = 1$ for all $n$. In many other situations, large flows are considered to be more important than small flows, then we want $w(n)$ be an increasing function in $n$. The weight function used in this thesis is presented and explained as follows.

First, assume that the cardinality of a randomly chosen flow can be modeled as a random variable $N$ with pmf $p_N(n) \triangleq \Pr(N = n)$. We use the following weight function:
\begin{equation} \label{eq:weight_function}
    w(n) = \frac{1}{n \cdot p_N(n)}.
\end{equation}
The motivation for using this weight function is explained here. The integral \[\int_{m_1}^{m_2} \! w(n) p_N(n)  \, \dd n\] is an approximation of the total weight of flows whose cardinalities are in the range $[m_1, m_2)$. With the weight function in (\ref{eq:weight_function}), we have $w(n) p_N(n) = \frac{1}{n}$; it is easy to verify that in this case the integral \[\int_{n^*}^{n^*(1+\epsilon)} \! w(n) p_N(n)  \, \dd n\] for $\epsilon > 0$ is independent of $n^*$. If we plot the curve $w(n)p_N(n)$ as a function of $n$ on log scale of $n$, then the area under the curve should be approximately the same in each decade interval: $[1,10)$, $[10,100)$, $[100,1000)$, $[1000,10000)$, etc. In another word, by choosing such a weight function, we put approximately the same total weight to the aggregate of flow cardinalities in each of these intervals.

\subsection{Zipf model} \label{subsec:zipf_model}

The weight function in (\ref{eq:weight_function}) can be applied to any given flow cardinality distribution. The particular distribution used for simulation in this thesis is the Zipf distribution. A random variable $N$ following the $\Zipf(\pi, n_{\max})$ distribution has the following pmf:
\begin{equation} \label{eq:zipf_pmf}
    p_N(n) = \frac{n^{-\pi}}{C}, \quad n \in \{1,2,\cdots,n_{\max}\},
\end{equation}
where $n_{\max}$ is an upper bound of the flow cardinality, $\pi$ is a parameter that controls the shape of the Zipf distribution ($\pi > 0$) and $C$ is a constant such that $\sum_{n=1}^{n_{\max}} p_N(n) = 1$.  

\begin{figure}[!h]
    \centering
    \includegraphics[scale=0.6]{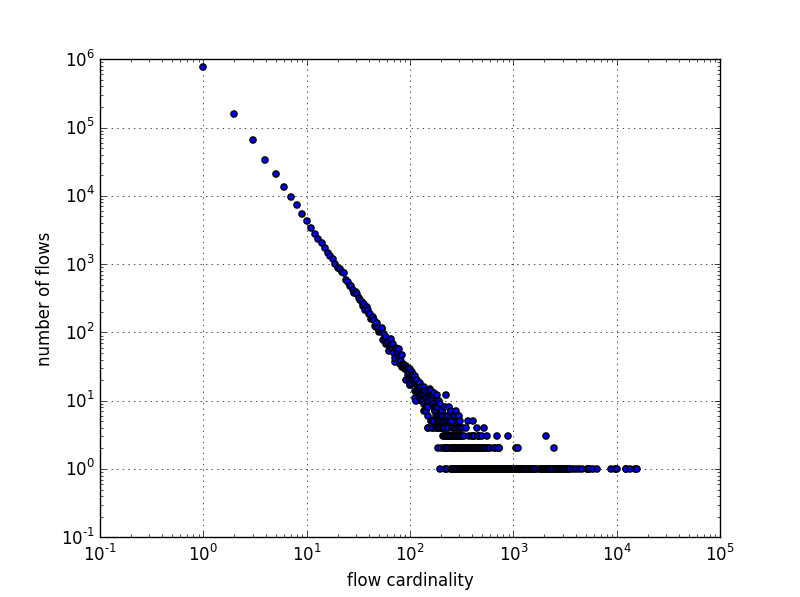}
    \caption{Distribution of simulated flow cardinalities according to $\Zipf(2.25, 10^5)$. Total number of flows: $1,116,535$. Average flow cardinality: $2.88$.}
    \label{fig:Zipf_2_25}
\end{figure}

The adoption of the Zipf model is motivated by the fact that Zipf's law underlies many Internet applications (see \cite{AdamicHuberman2002}). In Figure \ref{fig:Zipf_2_25}, we plot the distribution of a large number of simulated flow cardinalities, each generated independently according to a $\Zipf(2.25, 10^5)$ model. The validity of the model can be verified by observing the resemblance between Figure \ref{fig:Zipf_2_25} (from Zipf model) and Figure \ref{fig:trace_flow_cardi_dist} (from raw Internet data).

With $N \sim \Zipf(\pi, n_{\max})$, by (\ref{eq:weight_function}) and (\ref{eq:zipf_pmf}), we have 
\begin{equation}
    w(n) = C n^{\pi -1},
\end{equation}
where the constant $C$ can be omitted (set to 1) because when we compare the weighted square errors of two estimators, we only care about their relative values, which are unaffected by a constant factor. 

We remark that the specific estimators we will discuss in Chapters \ref{ch:generalize_theta} and \ref{ch:mle} do not reply on any particular distribution model. The Zipf model here is for two purposes: to complete the definition of the weighted square error for the performance evaluation of estimators and to generate random flow cardinalities for simulation; both purposes are independent of the estimation process.

Finally, we make a note on experimental evaluation of per-flow estimators. Given the close match between the real trace data and simulated trace data with a Zipf distribution, we will evaluate the estimators using simulated trace data. We generated $100$ simulated trace files. Each trace file contains $10^6$ flows, with the cardinality of each flow randomly and independently generated according to a $\Zipf(2.25, 10^5)$ distribution. A flow is presented in the file as a collection of packets with the same source address and distinct destination addresses; different flows have different source addresses (flow IDs). For one experiment, we process one such trace file and estimate the cardinalities of all the $10^6$ flows in it. For a given estimator, we perform $100$ independent experiments and evaluate its performance based on the statistics averaged over the $100$ experiments. All experiments are performed with $m = 200,000$ and $k = 512$. If each register uses $5$ bits, this setting uses $10^6$ bits to measure the cardinalities of $10^6$ flows, leading to one bit per flow on average.

\chapter{Virtual LogLog Estimator with Parameter $\theta$} \label{ch:generalize_theta}

The sketching process of the virtual LogLog algorithm has been described in Section \ref{subsec:vLL_sketching}. After the sketching process, we can offload the register values from the network measuring device for off-line query. The following per-flow estimation problem is to be solved:
\begin{center}
    \textbf{Given}: Register values $R[0], R[1], \cdots, R[m-1]$ and any flow's ID $f$. \\
    \textbf{Objective}: Estimate $n_f$, the number of distinct elements in flow $f$.
\end{center}
In this chapter, we first introduce $\LL_\theta$, a family of generalized LogLog estimators parameterized by $\theta$, for single-flow estimation. Then, with the similar idea from $\LL_\theta$, we propose $\vLL_\theta$, a family of generalized virtual LogLog estimators parameterized by $\theta$, for per-flow estimation. 

\section{The $\LL_\theta$ estimator for single-flow estimation} \label{sec:ll_theta}

The $\LL_\theta$ estimator is a class of estimators parameterized by $\theta$, which unifies and generalizes the LogLog and HLL estimators described in Section \ref{subsec:ll_hll}.  

\subsection{Generalized mean}

We start with the concept of \emph{generalized mean}. Given $k$ positive numbers $x_1, x_2, \cdots, x_k$, their \emph{generalized mean parameterized by $\theta$} is defined for nonzero $\theta$ by
\begin{equation}
    A_{\theta}(x_1, \cdots, x_k) = \left(\frac{1}{k} \sum_{i=1}^k x_i^\theta \right)^{\frac{1}{\theta}}, \quad  \theta \in \sR, \theta \not= 0.
\end{equation}
In the case of $\theta = 0$, we let
\begin{equation}
    A_0(x_1, \cdots, x_k) = \left(\prod_{i=1}^k x_i\right)^{\frac{1}{k}},
\end{equation}
which is in fact the limit of $A_\theta(x_1, \cdots, x_k)$ as $\theta \rightarrow 0$. Note that 
\begin{equation}
    A_\theta (x_1, \cdots, x_k) = \left\{\begin{array}{cccc}
        \min\{x_1, \cdots, x_k\} & if & \theta = -\infty \\
        \frac{k}{\frac{1}{x_1} + \cdots + \frac{1}{x_k}} & if & \theta = -1 \\
        \frac{x_1 + \cdots + x_k}{k} & if & \theta = 1 \\
        \max\{x_1, \cdots, x_k\} & if & \theta = \infty . \\
    \end{array} \right. 
\end{equation}
$A_\theta(x_1, \cdots, x_k)$ is commonly known as the \emph{arithmetic mean}, \emph{geometric mean}, and \emph{harmonic mean} of the $k$ numbers when $\theta = 1, 0, -1$, respectively. A notable property of the generalized mean function is that $A_\theta(x_1, \cdots, x_k)$ with a lesser $\theta$ is more robust to abnormally high values in obtaining the mean. Consider an example: for numbers $1,1,1,1,100$, $A_1 = 20.8$, $A_0 \approx 2.51$, $A_{-1} \approx 1.25$ and $A_{-\infty} = 1$.

\subsection{Unification and generalization of LogLog and HLL estimators}

Recall from Section \ref{subsec:ll_hll}, the LogLog and HLL estimators use the geometric mean and harmonic mean, respectively. Based on the generalized mean notation, we attempt to unify these two estimators by proposing the $\LL_\theta$ estimator:
\begin{equation} \label{eq:ll_theta_estimator}
    \widehat{n} = \LL_\theta(S) \triangleq \xi k A_\theta\left(2^{S[0]}, \cdots, 2^{S[k-1]}\right),
\end{equation}
where $\xi$ is a suitable coefficient. 

It is desirable if there exists a value of $\xi$ to make the estimator approximately unbiased and for which we can identify such value. Specifically, $\xi$ should not depend on $n$. We already know that for $\theta = 0$ (i.e. LogLog \cite{DurandFlajolet2003}) and $\theta = -1$ (i.e. HLL \cite{Flajoletetc2007}) such values of $\xi$ exist (by letting $\xi = \eta$ and $\xi = \gamma$ respectively). But we do not know if it is true for other values of $\theta$. The analysis of this estimator for general $\theta$ is difficult.\footnote{Interested readers may refer to the analyses of LogLog \cite{DurandFlajolet2003} and HLL \cite{Flajoletetc2007} algorithms to get an idea of the techniques used for this kind of analysis.} We resort to simulations to explore empirical evidence of the existence of this coefficient $\xi$.

\begin{figure}[!h]
\centering
\begin{subfigure}{0.5\textwidth}
  \centering
  \includegraphics[scale=0.45]{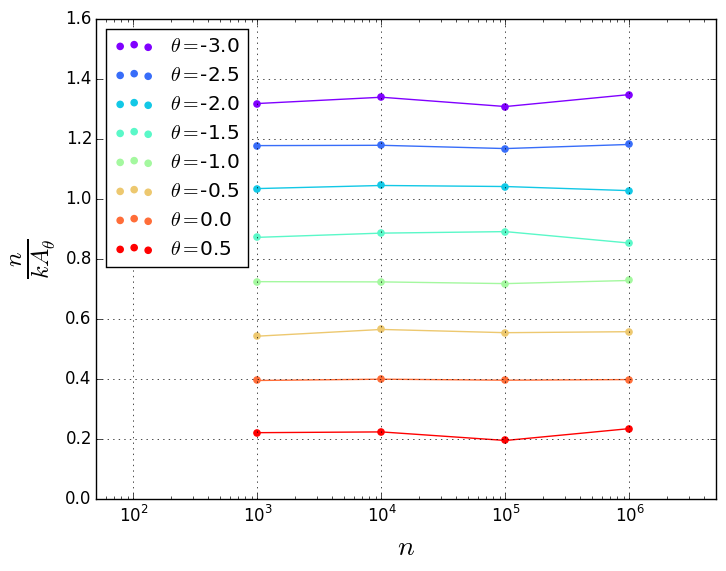}
  \caption{$k = 256$}
\end{subfigure}%
\begin{subfigure}{0.5\textwidth}
  \centering
  \includegraphics[scale=0.45]{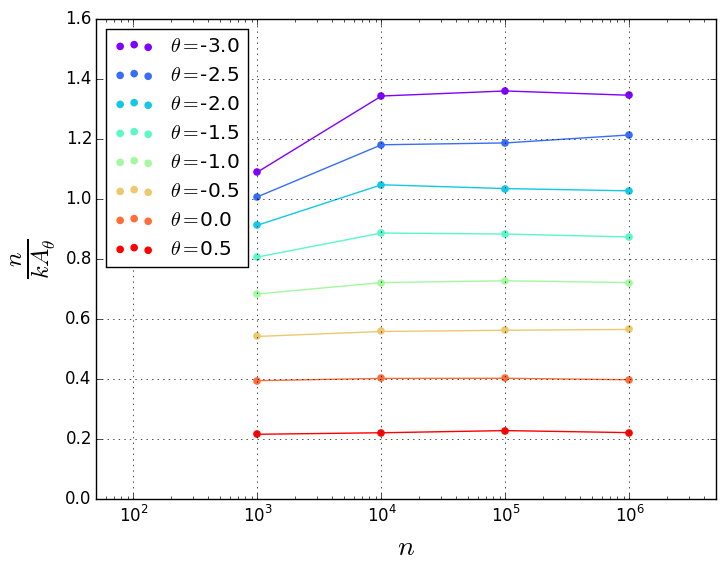}
  \caption{$k = 512$}
\end{subfigure}
\begin{subfigure}{0.5\textwidth}
  \centering
  \includegraphics[scale=0.45]{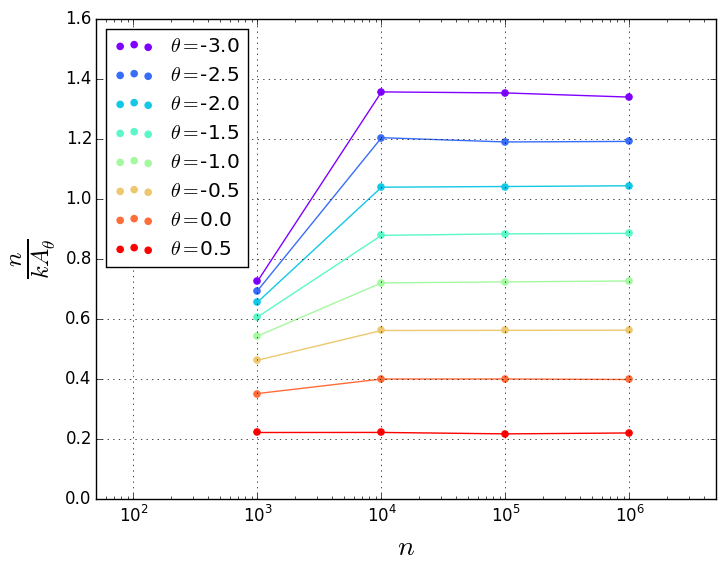}
  \caption{$k = 1024$}
\end{subfigure}
\caption{Empirical values of $\frac{n}{kA_\theta}$ for different $k$, $n$ and $\theta$'s.}
\label{fig:xi_theta_vs_n}
\end{figure}

Let $A_\theta$ denote $A_\theta\left(2^{S[0]},\cdots,2^{S[k-1]}\right)$ for short. In Figure \ref{fig:xi_theta_vs_n} we show empirical values of the ratio $\frac{n}{kA_\theta}$  for selected values of $k$, $n$ and $\theta$. Each dot in each of the sub-figures is generated by taking the average of $50$ independent experiment results. Experiments are performed on values of $\theta$ in $[-3.0, -2.9, -2.8, \dots, 0.9, 1.0]$, but plots are only shown for selected values of $\theta$ for better graph layout; plots for other values of $\theta$ have similar shapes and are at their expected positions in the figure.  

From the plots we see:
\begin{itemize}
 	\item For the same value of $\theta$ and $k$, $\frac{n}{kA_\theta}$ is almost a constant when $n$ gets large (say $n > 10000$). 
 	\item For the same value of $\theta$ and $n$, $\frac{n}{kA_\theta}$ is almost a constant for large $k$ ($256, 512$ or $1024$). 
\end{itemize}

We conclude that, at least for $\theta \in [-3.0, 1.0]$, there exists a value for the aforementioned coefficient $\xi$ that approximately depends only on $\theta$ for large $n$ and $k$. To reflect this dependence of $\xi$ on $\theta$, we denote this coefficient as $\xi_\theta$ instead. 

\begin{figure}[!h]
\centering
\includegraphics[scale=0.6]{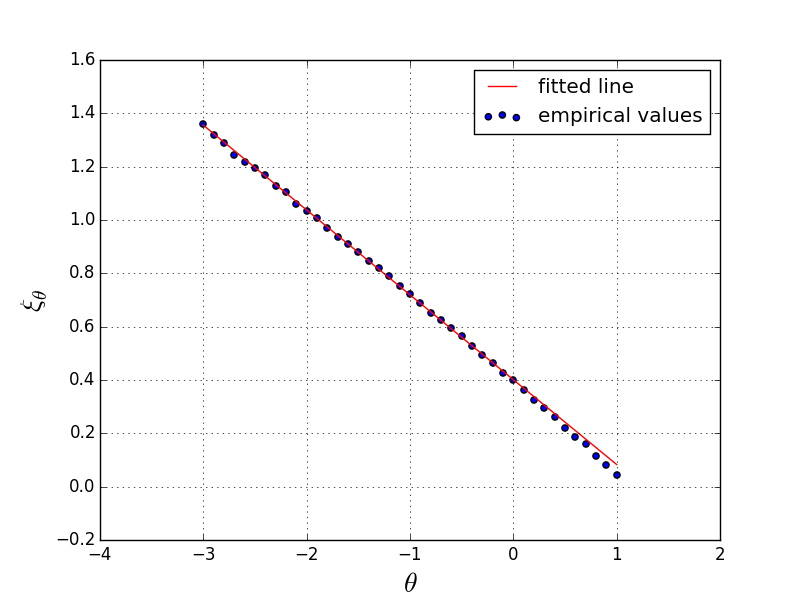}
\caption{Empirical values of $\xi_\theta$ as a function of $\theta$ and a fitted line showing the approximate linear relationship.}
\label{fig:xi_vs_theta}
\end{figure}

To investigate the relationship between $\xi_\theta$ and $\theta$, we plot values of $\xi_\theta$ computed empirically for selected values of $\theta$ in $[-3.0, 1.0]$, shown in Figure \ref{fig:xi_vs_theta}. Here, the values are generated empirically with fixed $n = 10^5$ and $k = 512$.

We observe an approximately linear relationship between $\xi_\theta$ and $\theta$ by
\begin{equation} \label{eq:xi_vs_theta_linear}
    \xi_\theta = 0.401 - 0.318\theta, \quad -3.0 \leq \theta \leq 1.0,
\end{equation}
which is plotted as the red line in Figure \ref{fig:xi_vs_theta}. The $\LL_\theta$ estimator is now completely defined by replacing $\xi$ in (\ref{eq:ll_theta_estimator}) with $\xi_\theta$ specified in (\ref{eq:xi_vs_theta_linear}). The claim that this family of estimators unifies the LogLog and HLL estimators can be verified by checking that $\xi_0 \approx \eta$ (for LogLog) and $\xi_{-1} \approx \gamma$ for (HLL). 

\begin{figure}[!h]
	\centering
	\includegraphics[scale=0.6]{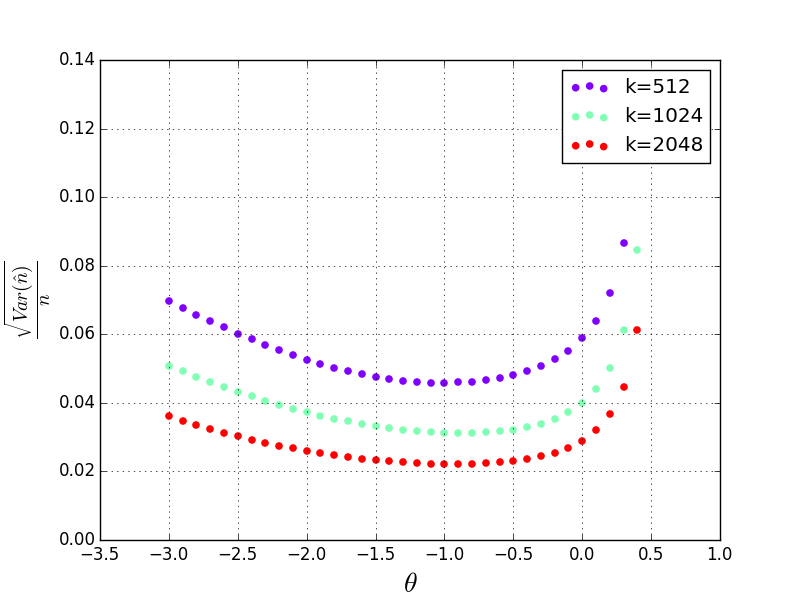}
	\caption{Empirically calculated relative standard error $\triangleq \frac{\sqrt{\Var(\widehat{n})}}{n}$ of the $\LL_\theta$ estimator vs. $\theta$ for selected $k$'s.}
	\label{fig:std_err_vs_theta_single_flow}
\end{figure}

A natural question one may ask next is if there exists an optimal value of $\theta$ for the $\LL_\theta$ estimator. To answer this question, we empirically calculate the relative standard error of the $\LL_\theta$ estimator for different values of $\theta$ (and for selected $k$'s). The results are plotted in Figure \ref{fig:std_err_vs_theta_single_flow}. 

According to the experiment results, for each fixed value of $k$, the optimal value of $\theta$ for the $\LL_\theta$ estimator is either $-1$ or $-0.9$; in the cases where $\LL_{-0.9}$ is the optimal, the difference between $\LL_{-1}$ and $\LL_{-0.9}$, in terms of relative standard error, is negligible. Recall the $\LL_{-1}$ estimator is the same as the HLL estimator. This verifies the claim that the HLL is near-optimal \cite{Flajoletetc2007}, which we briefly described in Section \ref{subsec:ll_hll}.

\section{The $\vLL_\theta$ estimator for per-flow estimation} \label{sec:vLL_theta}

With the $\LL_\theta$ estimator defined, we now introduce the $\vLL_\theta$ estimator for per-flow estimation. We start with a high-level idea of the estimator and the motivation for introducing the parameter $\theta$ here. 

Just like for single-flow estimation where we infer the cardinality of a flow from its register array, for per-flow estimation we can infer the cardinality of a given flow $f$ from its \emph{virtual} register array $R_f$. We can directly apply the $\LL_\theta$ estimator on $R_f$ to give an estimate of the total number of distinct elements distributed to $R_f$. But this estimate involves the noise brought by other flows that have registers shared with flow $f$; so we probably need a different bias correction coefficient to calibrate this rough estimate. While for the single-flow case $\theta = -1$ is near-optimal, it is not immediately clear whether $\theta = -1$ with a suitable bias correction coefficient is still near-optimal in the per-flow case. 

Recall that $A_\theta$ with a lesser value of $\theta$ is more robust to abnormally high values in obtaining the mean. For per-flow estimation, due to register sharing, large flows can cause much noise to some small flows by causing abnormally high register values. Therefore, we speculate that an estimator with a lesser value of $\theta$ might work better in the per-flow case due to its robustness against bursty noise from large flows. This motivates our investigation of the $\vLL_\theta$ estimator, introduced in the rest of this section.

\subsection{Review of vHLL estimator and the generalization to $\vLL_\theta$ estimator} \label{subsec:review_vHLL}

We start with a review of the virtual HyperLogLog (vHLL) estimator introduced in \cite{Xiaoetc2015}. Let $n_T$ be the total aggregate cardinality of all flows, i.e. the sum of the cardinalities of all flows in the packet stream. Let $n_{f^+}$ be the total number of distinct elements distributed to the register array $R_f$, which include the distinct elements from flow $f$ and those from other flows (we call noise elements). Suggested in \cite{Xiaoetc2015} is the following approximate relationship between $n_f, n_{f^+}$ and $n_T$:
\begin{equation} \label{eq:vHLL_original_idea}
    n_{f^+} - n_f \approx \frac{k}{m}(n_T - n_f),
\end{equation}
which can be interpreted as: the number of noise elements received by $R_f$ is the total number of elements from flows other than $f$ scaled by $\frac{k}{m}$ --- the ratio of the number of registers in $R_f$ to the number of registers in $R$. The assumption here is that noise elements are roughly uniformly distributed to all the $m$ registers in the pool, which is a fair approximation when the number of flows and the number of registers for each flow are both sufficiently large \cite{Xiaoetc2015}. Rearranging (\ref{eq:vHLL_original_idea}) gives us 
\begin{equation}
    n_f \approx  - \frac{k}{m-k}n_T + \frac{m}{m-k} n_{f^+}.
\end{equation}
Now the values of $n_T$ and $n_{f^+}$ are not directly available so we need their estimates $\widehat{n}_T$ and $\widehat{n}_{f^+}$. It has been shown in Section 6.1 of \cite{Xiaoetc2015} that if $\widehat{n}_{f^+}$ and $\widehat{n}_T$ are close to $n_f + \frac{k}{m}(n_T - n_f)$ (which approximately equals to $n_{f^+}$ by (\ref{eq:vHLL_original_idea})) and $n_T$ respectively, then $\widehat{n}_f$ is an approximately unbiased estimator of $n_f$. 

There are two possible methods to estimate $n_T$. The first method is to treat $n_T$ as the cardinality of a \emph{grand flow} -- the flow containing all the distinct packets in the stream. In another word, if a packet is abstracted as a $(f,x)$ pair, to estimate $n_T$ is to estimate the total number of distinct $f-x$ pairs (e.g. source-destination pair) in the stream. This is a single-flow cardinality estimation problem and, as discussed in Section \ref{sec:single_flow_estimation}, can be solved using the HLL algorithm with an additional few hundreds of bytes, which is negligible compared to the memory for main register pool $R$. The second method is to use $\widehat{n}_T = \HLL(R)$ as a rough estimator, based on the assumption that all the elements of the grand flow are distributed approximately uniformly over $R$. Either method works fine in practice. So $\widehat{n}_T$ is relatively easy to obtain. 

Now, for $n_{f^+}$, the vHLL estimator applies the HLL single-flow estimator on $R_f$ to obtain an estimate, i.e.
\begin{equation} \label{eq:HLL_Rf}
    \widehat{n}_{f^+} = \HLL(R_f) = \xi_{-1} k A_{-1} \left(2^{-R_f[0]}, \cdots,  2^{-R_f[k-1]}\right).
\end{equation}
 
Combining the ideas above, the vHLL estimator for flow $f$ is summarized as:
\begin{equation} \label{eq:vHLL_estimator}
    \widehat{n}_f =  - \frac{k}{m-k}\widehat{n}_T + \frac{m}{m-k} \xi_{-1} k A_{-1} \left(2^{-R_f[0]}, \cdots,  2^{-R_f[k-1]}\right).
\end{equation}

We consider a generalization of the vHLL estimator by replacing the above harmonic mean $A_{-1}$ with a generalized mean $A_\theta$. More specifically, given any value $\theta$, let 
\begin{equation}
\widetilde{n}_f(\theta) = k \cdot A_\theta\left(2^{R_f[0]}, \cdots, 2^{R_f[k-1]}\right)
\end{equation}
be a rough estimate of $n_f$. Then we calibrate this rough estimate by suitable additive and multiplicative constants $\alpha_\theta$ and $\beta_\theta$ to obtain a better estimate, i.e.
\begin{equation} \label{eq:vLL_theta_estimator_basic}
\widehat{n}_f(\theta) = \alpha_\theta + \beta_\theta\widetilde{n}_f(\theta).
\end{equation}
We discuss how to set the values of $\alpha_\theta$ and $\beta_\theta$ in Section \ref{subsec:alpha_and_beta_theta}.  
 
\subsection{How to set $\alpha_\theta$ and $\beta_\theta$} \label{subsec:alpha_and_beta_theta}

For a given $\theta$, one way to set the values of $\alpha_\theta$ and $\beta_\theta$ is through \emph{empirical error minimization}, explained as follows. Recall from the end of Section \ref{subsec:zipf_model}, we have 100 simulated trace files, each can be considered as a training sample. Suppose that in one training sample we have $M$ flows with cardinalities $n_1, n_2, \cdots, n_M$, and by the $\vLL_\theta$ estimator specified in (\ref{eq:vLL_theta_estimator_basic}) we obtain corresponding estimates $\widehat{n}_1(\theta), \widehat{n}_2(\theta), \cdots, \widehat{n}_M(\theta)$. Then we can find values of $\alpha_\theta$ and $\beta_\theta$ that minimize the \emph{weighted square error} defined in Section \ref{sec:wse} on this training sample. That is 
\begin{equation}
(\alpha_\theta, \beta_\theta) = \arg \min_{(\alpha,\beta)} \sum_{i=1}^M \left(n_i - \widehat{n}_i(\theta)\right)^2 w(n_i) = \arg \min_{(\alpha,\beta)} \sum_{i=1}^M \left(n_i - (\alpha + \beta \widetilde{n}_i(\theta))\right)^2 w(n_i),
\end{equation}
where we let $w(n_i) = n_i^{\pi-1}$ and $\pi = 2.25$. The solution to the above minimization problem is standard:
\begin{equation} \label{eq:linear_regression_solution}
\beta_\theta = \frac{\overline{xy} - \overline{x} \cdot \overline{y}}{\overline{x^2} - \overline{x}^2}, \quad \alpha_\theta = \overline{y} - \beta_\theta \overline{x},
\end{equation}
where
\begin{equation*}
\overline{x} = \frac{\sum_{i=1}^M w(n_i) \widetilde{n}_i(\theta)}{\sum_{i=1}^M w(n_i)},   \overline{y} = \frac{\sum_{i=1}^M w(n_i) n_i}{\sum_{i=1}^M w(n_i)}, \overline{xy} = \frac{\sum_{i=1}^M w(n_i) \widetilde{n}_i(\theta) n_i}{\sum_{i=1}^M w(n_i)}, \overline{x^2} = \frac{\sum_{i=1}^M w(n_i) \widetilde{n}_i^2(\theta)}{\sum_{i=1}^M w(n_i)}.
\end{equation*}
Since we have 100 training samples, we can obtain the values of $\alpha_\theta$ and $\beta_\theta$ for each of the samples using the above method and use their average values for the estimator. The values of $\alpha_\theta$ and $\beta_\theta$ obtained in this way are optimal in the sense that they minimize the weighted square error of the training samples (i.e. the \emph{empirical error}). The drawback of this approach is obvious. It only works well for the training samples generated according to a specific Zipf distribution; there is no performance guarantee for other data sets. Also, it relies on the specific definition of the weighted square error (including the weight function and the cardinality distribution model), which may not be universal for all applications.

We suggest the following alternative and practical choice of $\alpha_\theta$ and $\beta_\theta$ for the $\vLL_\theta$ estimator, which does not reply on any assumption on weight function or cardinality distribution model:
\begin{equation}
\beta_\theta = \frac{m}{m-k}\xi_\theta \quad \text{and} \quad \alpha_\theta = -\frac{k}{m-k} \widehat{n}_T,
\end{equation}
where $\xi_\theta$ can be calculated using (\ref{eq:xi_vs_theta_linear}) and $\widehat{n}_T$ can be calculated using either of the two methods mentioned in Section \ref{subsec:review_vHLL} (e.g. $\widehat{n}_T = \HLL(R)$). Note that this is a direct generalization of the vHLL estimator in (\ref{eq:vHLL_estimator}). Also note that in this case $\alpha_\theta$ does not depend on $\theta$. 

\begin{figure}[!h]
	\centering
	\begin{subfigure}{0.5\textwidth}
		\centering
		\includegraphics[scale=0.4]{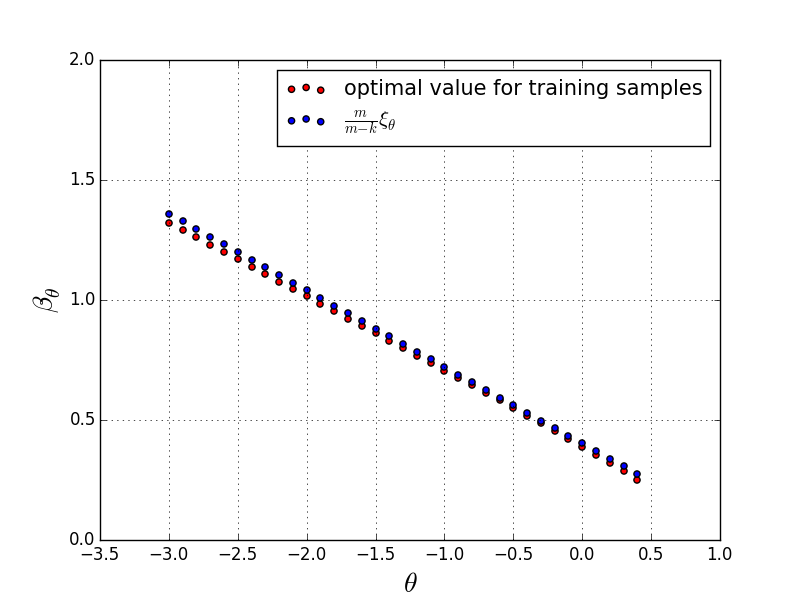}
		\caption{}
	\end{subfigure}%
	\begin{subfigure}{0.5\textwidth}
		\centering
		\includegraphics[scale=0.4]{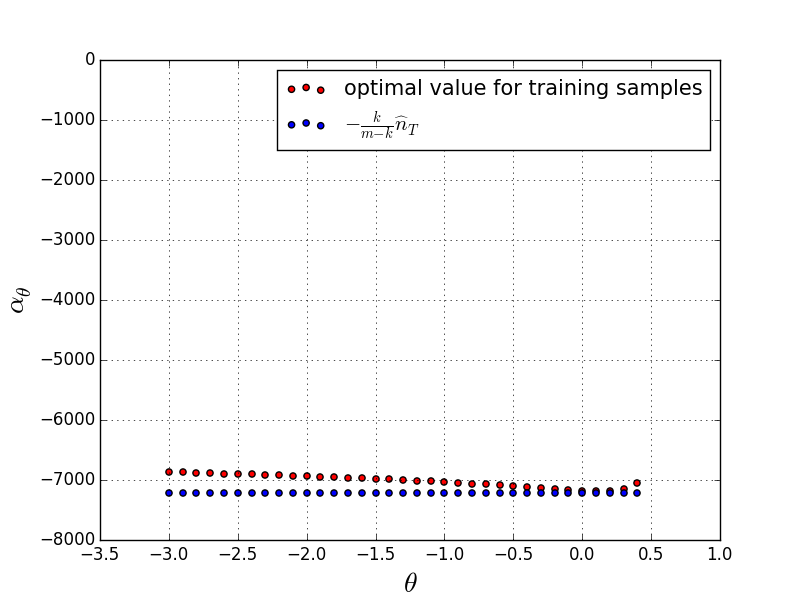}
		\caption{}
	\end{subfigure}
	\caption{The values of $\alpha_\theta$ and $\beta_\theta$ vs. $\theta$ by two different methods.}
	\label{fig:alpha_beta_vs_theta}
\end{figure}

In Figure \ref{fig:alpha_beta_vs_theta}, we plot the empirical values of $\alpha_\theta$ and $\beta_\theta$ obtained by these two different methods through experiments on the $100$ training samples. The plots show that our suggested values of $\alpha_\theta$ and $\beta_\theta$ are good as they are close to the optimal values of $\alpha_\theta$ and $\beta_\theta$ for these training samples. 

\subsection{Summary of the $\vLL_\theta$ estimator}

We summarize the $\vLL_\theta$ estimator here. For a given $\theta \in [-3.0, 1.0]$, the $\vLL_\theta$ estimator estimates the cardinality of a given flow $f$ by 

\begin{equation} \label{eq:vLL_theta_final}
\widehat{n}_f = -\frac{k}{m-k}\widehat{n}_T + \frac{m}{m-k}\xi_\theta k A_\theta\left(2^{R_f[0]}, \cdots, 2^{R_f[k-1]}\right),
\end{equation}
where 
\begin{equation}
\xi_\theta = 0.401 - 0.318\theta, \quad -3.0 \leq \theta \leq 1.0,
\end{equation}
and $\widehat{n}_T$ is an estimate of the total aggregate cardinality of all the flows and can be obtained by calculating $\HLL(R)$ in practice.

Note that it could happen that Equation (\ref{eq:vLL_theta_final}) produces an estimate $\widehat{n}_f < 1$, which is impossible in practice. In such case, we can simply reset $\widehat{n}_f = 1$.

\section{Experimental performance evaluation}

\begin{figure}[!h]
	\centering
	\includegraphics[scale=0.55]{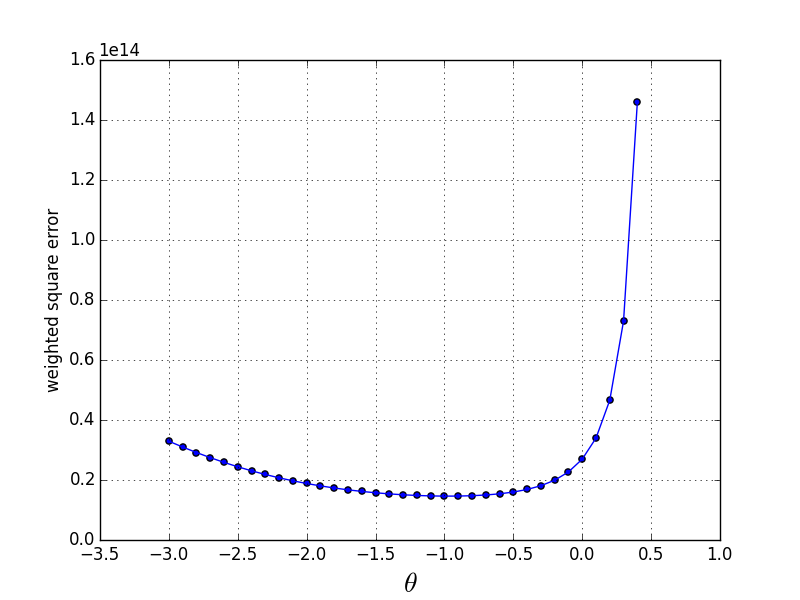}
	\caption{Experimental values of WSE($\vLL_\theta$) vs. $\theta$.}
	\label{fig:wse_vs_theta}
\end{figure}

We ran experiments on the 100 simulated trace files to evaluate the performance of the $\vLL_\theta$ estimator for different values of $\theta$, in terms of the weighted square error defined in Section \ref{sec:wse}. The results are plotted in Figure \ref{fig:wse_vs_theta}. Results for $\theta \geq 0.5$ are not plotted in the graph because the values are comparatively too large (which means the corresponding $\vLL_\theta$ estimators are bad). 

According to the experiment results, the $\vLL_{-1}$ estimator has the best performance, which refutes our speculation at the beginning of Section \ref{sec:vLL_theta} that a value of $\theta$ lesser than $-1$ might be optimal for the per-flow estimation case. 

A possible explanation of this phenomenon is as follows. With the weight function defined in Section \ref{sec:wse} and $\Zipf(2.25, 10^5)$ used for flow cardinality distribution, we have that $w(n) \propto n^{1.15}$. This means we put much larger weights on large flows compared to small flows. Therefore the weighted square error of an estimator is largely determined by how well the estimator estimates large flows. But the influence of bursty noise on large flows is much weaker than that on small flows; in another word, the signal-to-noise ratio for the cardinality estimation of a large flow is larger than that for small flows. Therefore, as we have discussed for the single-flow case where $\theta = -1$ is near-optimal, for per-flow case $\theta = -1$ is also likely to give good estimates for large flows and hence results in a smaller weighted square error.

\chapter{Maximum Likelihood Estimator} \label{ch:mle}

In this chapter we derive an alternative approach to the same per-flow estimation problem stated at the beginning of Chapter \ref{ch:generalize_theta}: a maximum-likelihood estimator.

\section{Formulation}

We model the values of the $k$ registers in $R_f$ after the sketching process as a random vector  
\begin{equation*}
    Z_f = \left(Z_{f,0}, Z_{f,1}, \cdots, Z_{f,k-1}\right),
\end{equation*}
where $Z_{f,i} \in \{0,\cdots,r_{\max}\}$. The $r_{\max}$ is the maximum value that can be stored in a register. For example, if the register has $5$ bits, then $r_{\max} = 2^5-1 = 31$. Register values $R_f = (R_f[0], \cdots, R_f[k-1])$ is an instance of the random vector $Z_f$. Let $p_n(z_0, \cdots, z_{k-1})$ be the pmf of $Z_f$ (i.e. the joint pmf of the $k$ random variables) given that $n_f = n$, i.e. it is the likelihood function of $n$:
\begin{equation}
    p_n(z_0, \cdots, z_{k-1}) \triangleq \Pr\left\{Z_f = (z_0, \cdots, z_{k-1}) \, \Big| \, n_f = n\right\}.
\end{equation}
The maximum likelihood estimator finds the value of $n$ that maximizes the likelihood function based on the instance of register values in $R_f$. An equivalent formulation is to maximize the natural log of the likelihood function 
\begin{equation} \label{eq:log_likelihood_function}
    L_f(n) \triangleq \ln p_n \left( R_f[0], \cdots, R_f[k-1] \right).
\end{equation}
The maximum-likelihood estimator of flow $f$'s cardinality is then given by
\begin{equation} \label{eq:mle}
    \widehat{n}_{f, \ML} = \arg \max_{n} L_f(n), \quad n \in \{1,2,\cdots,n_{\max}\},
\end{equation}
where $n_{\max}$ is an upper bound of the cardinality of a flow. 

\section{Derivation of the likelihood function}

In this section we derive an analytical expression for the likelihood function $p_n(z_0, \cdots, z_{k-1})$.

\subsection{Outline of derivation}

Suppose we have two copies of the register pool $R$: $R^{(1)}$ and $R^{(2)}$, both initialized with $0$'s. Consider the following process:
\begin{enumerate}
    \item Repeat the virtual LogLog sketching process described in Section \ref{subsec:vLL_sketching} for all elements, except those from flow $f$, with register pool $R^{(1)}$.
    \item Repeat the sketching process only for elements from flow $f$ with register pool $R^{(2)}$.
    \item Merge the register values in $R^{(1)}$ and $R^{(2)}$ by a register-wise max operation. 
\end{enumerate}
The register values in the merged pool as described above are exactly the same as those in $R$ after the original sketching process. This is because the register values at the end of the sketching process are not affected by the order of arrival of the elements in the stream. 

Let $R_f^{(1)}$ ($R_f^{(2)}$) denote the virtual register array for flow $f$ constructed by selecting registers from $R^{(1)}$ ($R^{(2)}$), whose indices in $R^{(1)}$ ($R^{(2)}$) are the same as those of $R_f$ in $R$. Then define the following random variables:
\begin{center}
    $Z' \triangleq$ the value of an arbitrary register in $R^{(1)}$.
\end{center}
\begin{center}
    $W_n \triangleq$ the value of an arbitrary register in $R_f^{(2)}$, conditioned on $n_f = n$.
\end{center}
\begin{center}
    $Z_n \triangleq$ the value of an arbitrary register in $R_f$, conditioned on $n_f = n$.
\end{center}
$Z'$ represents the background noise in our estimation caused by elements from flows other than $f$. $W_n$ represents the impact of elements from flow $f$ itself. We model the values of the $k$ corresponding registers in $R_f^{(1)}$ and $R_f^{(2)}$ also as random vectors, respectively, i.e.
\begin{align*}
    Z_f^{(1)} &= \left( Z_{f,0}^{(1)}, Z_{f,1}^{(1)}, \cdots, Z_{f,k-1}^{(1)} \right), \\
    Z_f^{(2)} &= \left( Z_{f,0}^{(2)}, Z_{f,1}^{(2)}, \cdots, Z_{f,k-1}^{(2)} \right).
\end{align*}
Then we have 
\begin{align*}
    Z_{f,0}^{(1)}, Z_{f,1}^{(1)}, \cdots, Z_{f,k-1}^{(1)} &\sim Z', \\
    Z_{f,0}^{(2)}, Z_{f,1}^{(2)}, \cdots, Z_{f,k-1}^{(2)} &\sim W_n, \\
    Z_{f,0}, Z_{f,1}, \cdots, Z_{f,k-1} &\sim Z_n, \\
    Z_{f,i} = \max\{Z_{f,i}^{(1)}, Z_{f,i}^{(2)}\}, &\quad i = 0, 1, \cdots, k-1.
\end{align*}
Therefore 
\begin{equation} \label{eq:Zn}
    Z_n = \max\{Z', W_n\}.
\end{equation}
Here is an outline of the derivation of an analytical expression for $p_n(z_0, \cdots, z_{k-1})$:
\begin{enumerate}
	\item Find the distribution of $Z'$ and argue that $Z_{f,0}^{(1)}, Z_{f,1}^{(1)}, \cdots, Z_{f,k-1}^{(1)}$ are independent. We will see that it is difficult to obtain the exact distribution of $Z'$, but there exists a convenient and close approximation to it.
	\item Find the distribution of $W_n$ and argue that $Z_{f,0}^{(2)}, Z_{f,1}^{(2)}, \cdots, Z_{f,k-1}^{(2)}$ are independent under the assumption that $n$ is large. 
    \item With the distribution of $Z'$ and $W_n$, find the distribution of $Z_n$ by (\ref{eq:Zn}). 
    \item With the independence of $Z_{f,0}^{(1)}, Z_{f,1}^{(1)}, \cdots, Z_{f,k-1}^{(1)}$ and $Z_{f,0}^{(2)}, Z_{f,1}^{(2)}, \cdots, Z_{f,k-1}^{(2)}$, we have that $Z_{f,0}, Z_{f,1}, \cdots, Z_{f,k-1}$ are independent. Then we can factor $p_n(z_0,\cdots,z_{k-1})$ out as the product of $k$ pmfs of a single variable $Z_n$. 
\end{enumerate}

\subsection{Distribution of $Z'$ and independence of $Z_{f,0}^{(1)}, \cdots, Z_{f,k-1}^{(1)}$}

One possible approach to obtain an approximate expression of the distribution of $Z'$ is as follows. First assume that the total number of flows (excluding flow $f$) is known and a flow's cardinality can be modeled as a random variable (e.g. Zipf). Then we can model the number of distinct elements distributed to an arbitrary register by a random variable $Y$ and calculate the distribution of $Y$ by adding up the influences from all the individual independent flows on that register. This calculation involves a high-dimensional convolution, which is computationally costly. We can perhaps avoid the convolution by appealing to the central limit theory for approximation. However this approximation gives us a continuous distribution for $Y$, from which we need to derive the CDF of $Z'$ that only takes discrete values (and really concentrates on only a few values as we shall see soon). This calculation may be highly inaccurate and is somewhat complicated. Moreover, in practice we usually do not know the total number of flows beforehand and it is undesirable that our estimator relies on any particular cardinality distribution model. 

Another approach to approximate the distribution of $Z'$, which circumvents the above difficulties and complications, is the following. Define a random variable $Z$:
\begin{center}
    $Z \triangleq$  the value of an arbitrary register in $R$.
\end{center} 
We claim that, for any practical purpose, the distribution of $Z'$ can be directly approximated by the distribution of $Z$, i.e. 
\begin{equation} \label{eq:ZdashequalZ}
    F_{Z'}(i) \approx F_{Z}(i), \quad 0 \leq i \leq r_{\max},
\end{equation}
where $F_{Z'}$ and $F_Z$ are the CDFs of $Z'$ and $Z$, respectively. This approximation is based on the following two considerations:
\begin{itemize}
    \item Register values in $R^{(1)}$ differ from those in $R$ at no more than $k$ registers, because flow $f$ can only affect $k$ registers. If $k \ll m$, the difference in the CDF of $Z'$ compared to that of $Z$ is small.
    \item Assume there are many flows in the stream and $n_f$ is small compared to the total cardinalities of all other flows. As packets are being distributed to a register, it becomes harder and harder for the register's value to further increase, because it requires an much less likely (with geometrically decaying probability) hashed value to occur.
\end{itemize}

\begin{figure}[H]
\centering

\includegraphics[scale=0.6]{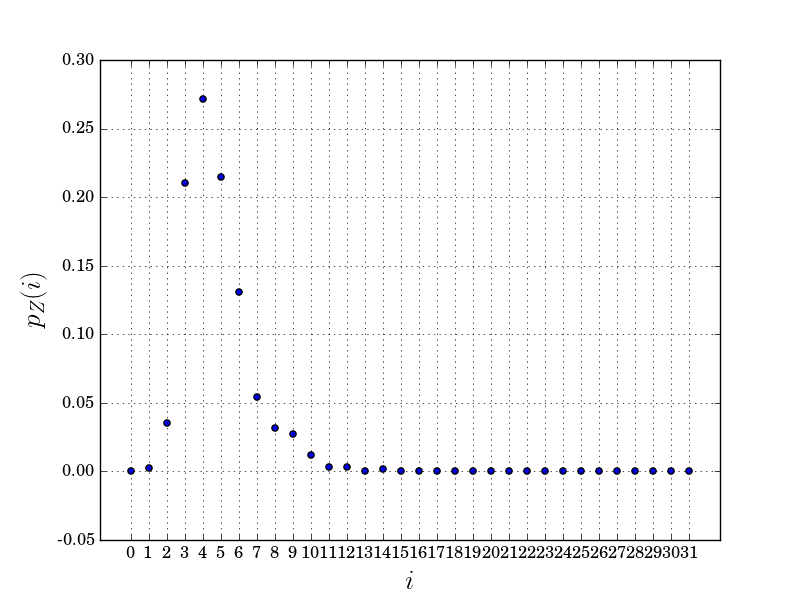}

\includegraphics[scale=0.6]{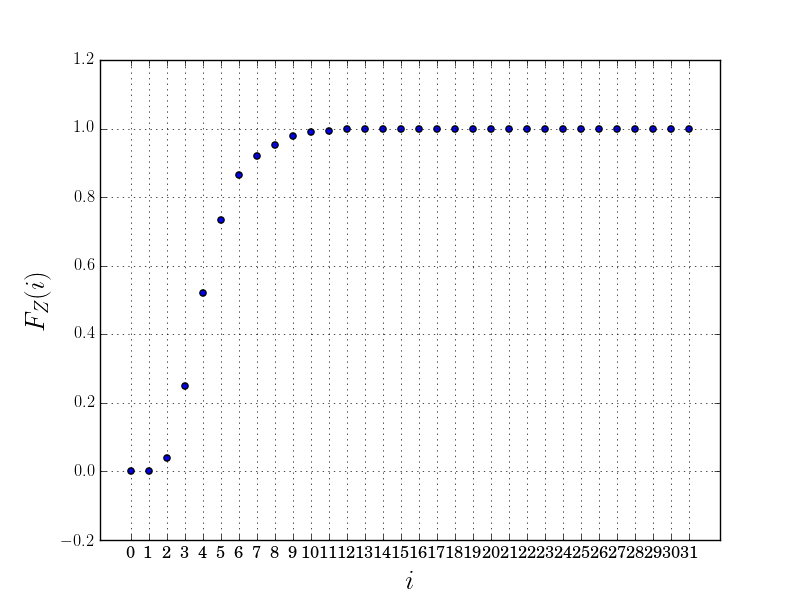}

\caption{Sample distribution of $Z$ generated using real trace data. Top: pmf of $Z$. Bottom: CDF of $Z$. In this case, $m = 200,000$ and $k = 512$.}
\label{fig:sample_Z_dist}
\end{figure}

The analytical distribution of $Z$ is also difficult to find. However, we can obtain an empirical distribution of $Z$ directly from the register values in the pool at the end of the sketching process:
\begin{equation} \label{eq:Z_cdf}
    F_Z(i) \triangleq \Pr(Z \leq i) \approx \frac{\displaystyle \sum_{j=0}^{m-1} \ind_{\{R[j] \leq i\}}}{m}, \; 0 \leq i \leq r_{\max} .
\end{equation}
In Figure \ref{fig:sample_Z_dist} we show a sample distribution of $Z$ generated using real trace data (the same data used for plotting Figure \ref{fig:trace_flow_cardi_dist}). From the figure we can see that the values of most of the registers are centered around $2$ to $7$.

Since $Z'$ is the distribution of an arbitrary register value in $R^{(1)}$, including the $k$ registers in $R_f^{(2)}$. We have $Z_{f,0}^{(1)}, Z_{f,1}^{(1)}, \cdots, Z_{f,k-1}^{(1)} \sim Z'$. Since the $k$ registers are randomly selected from the pool with replacement, their independence is guaranteed by a good choice of the hash function $G$.

\subsection{Distribution of $W_n$ and independence of $Z_{f,0}^{(2)}, \cdots, Z_{f,k-1}^{(2)}$}

To find the distribution of $W_n$, we first define a random variable $X_n$:
\begin{center}
    $X_n \triangleq$ the number of distinct elements distributed to an \\ arbitrary register in $R_f^{(2)}$, conditioned on $n_f = n$. 
\end{center}
It is easy to see $X_n \sim \Binom \left(n, \frac{1}{k}\right)$. The number of distinct elements distributed to each of the $k$ registers in $R_f^{(2)}$ has the same distribution as $X_n$, but they are not independent of each other (because they sum to $n$). However, if $n$ is large, we can approximate $n$ by a Poisson random variable with mean $n$. Under this Poissonization approximation, $X_n \sim \Poisson\left(\frac{n}{k}\right)$ and the number of distinct elements distributed to each of the $k$ registers in $R_f^{(2)}$ will then be independent (see Chapter 5.4 of \cite{MitzenmacherUpfal} for this Poissonization trick). 

As discussed in Section \ref{subsec:ll_hll}, $W_n$ can be regarded as the maximum of $X_n$ i.i.d. geometric random variables with parameter $1/2$. Given that $X_n \sim \Poisson\left(\frac{n}{k}\right)$, the CDF of $W_n$ can be calculated as follows. For $0 \leq i \leq r_{\max}$, 
\begin{align}
    \Pr\{W_n \leq i\} &= \sum_{j=0}^n \Pr(X_n = j) \Pr\{\text{the maximum of $j$ i.i.d. $\Geo(0.5)$ r.v.'s} \leq i\} \nonumber \\
    &= \sum_{j=0}^n \frac{(n/k)^j}{j!} e^{-n/k} \left(1- \frac{1}{2^i}\right)^j \nonumber \\
    &= \sum_{j=0}^n
         \frac{\left(\frac{n}{k}\left(1-\frac{1}{2^i}\right)\right)^j}{j!} e^{-n/k} \nonumber \\
    &= e^{-\frac{n}{k}\frac{1}{2^i}} \label{eq:Wn_CDF}.
\end{align}
Under the Poissonization approximation, $Z_{f,0}^{(2)}, Z_{f,1}^{(2)}, \cdots, Z_{f,k-1}^{(2)}$ are independent and have the same distribution with $W_n$. We emphasize that this independence follows from the assumption that $n$ is large.

\subsection{Distribution of $Z_n$ and independence of $Z_{f,0}, \cdots, Z_{f,k-1}$}

Recall that, by construction, $Z_n = \max\{Z', W_n\}$. Therefore the CDF of $Z_n$ is:
\begin{equation*}
    \Pr\{Z_n \leq i\} = \Pr\{\max\{Z',W_n\} \leq i\} =\Pr\{Z' \leq i\} \Pr\{W_n \leq i\} \approx F_{Z}(i)F_{W_n}(i),  \quad 0 \leq i \leq r_{\max},
\end{equation*}
where $F_Z(i)$ and $F_{W_n}(i)$ are the CDF of $Z$ and $W_n$ respectively. We have used that $Z'$ and $W_n$ are independent of each other and the distribution of $Z'$ can be approximated by that of $Z$.
Hence, the pmf of $Z_n$ can be approximated by
\begin{equation} \label{eq:Zn_pmf}
    p_{Z_n}(i) = \left\{\begin{array}{cccc}
        F_Z(i) F_{W_n}(i) & when & i = 0  \\
        F_Z(i) F_{W_n}(i) - F_Z(i-1) F_{W_n}(i-1) & when & 1 \leq i \leq r_{\max},
    \end{array} \right.
\end{equation}
where 
\begin{equation} \label{eq:Wn_cdf}
    F_{W_n}(i) = e^{-\frac{n}{k}\frac{1}{2^i}} \quad \text{with Poissonization of $n$}. \\
\end{equation}
We have argued the independence of $Z_{f,0}^{(1)}, \cdots, Z_{f,k-1}^{(1)}$ and the independence of $Z_{f,0}^{(2)}, \cdots, Z_{f,k-1}^{(2)}$ under certain conditions. But $Z_{f,i} = \max\left\{Z_{f,i}^{(1)}, Z_{f,i}^{(2)}\right\}$ and $Z_{f,i}^{(1)}$ is independent of $Z_{f,i}^{(2)}$ by construction. So $Z_{f,0}, \cdots, Z_{f,k-1}$ are also independent of each other under the same conditions. The consequence of this independence is that we can factor the $k$-variable likelihood function $p_n(z_0, \cdots, z_{k-1})$ into the product of $k$ single-variable pmfs, based on the distribution of $Z_n$. That is, under this approximation, the log likelihood function in (\ref{eq:log_likelihood_function}) can be re-written as
\begin{equation} \label{eq:log_likelihood_function2}
    L_f(n) =  \sum_{j=0}^{k-1} \ln p_{Z_n} \left(R_f[j] \right).
\end{equation}

\section{Implementation}

It would be nice if we could obtain a closed-form analytical expression of the value of $n$ that maximizes the log-likelihood function $L_f(n)$. Unfortunately it turns out to be difficult. In order to search for this value, we explore properties of $L_f(n)$ that might be helpful.

\subsection{Concavity of the log-likelihood function}

In this subsection we show that $L_f(n)$ has the decreasing increment property with respect to $n$. That is, $L_f(n)$ is the restriction of a concave function to integer values. A by-product of this property is that $L_f(n)$ must have a global maximum over the possible values of $n$. We will call this property ``concave" or ``concavity" for convenience henceforth. 

Suppose for now that the possible values of $n$ is a continuous range. First, since $R_f[j] \in \{0,\cdots,r_{\max}\}$, we have 
\begin{equation}
    L_f(n) = \sum_{j=0}^{k-1} \ln p_{Z_n} \left(R_f[j]\right) = \sum_{i = 0}^{r_{\max}} c_i \ln p_{Z_n}(i),
\end{equation}
for some non-negative integer constants $c_0, c_1, \cdots, c_{r_{\max}}$ such that $\sum_{i=1}^{r_{\max}} c_i = k$ (i.e. the constants represent the empirical pmf of $(R_f[0], \cdots, R_f[k-1])$). Therefore a sufficient condition for $L_f(n)$ to be concave in $n$ is that $\ln p_{Z_n}(i)$ is concave in $n$ for each fixed value $i \in \{0,\cdots,r_{\max}\}$. Recall the pmf of $Z_n$ from (\ref{eq:Zn_pmf}) and (\ref{eq:Wn_cdf}). Let us denote $a_i \triangleq F_Z(i)$ and $b_i \triangleq e^{-\frac{1}{k2^i}}$, so $a_i, b_i \geq 0$. Then we have
\begin{equation*}
    \ln p_{Z_n}(i) = \left\{\begin{array}{ccc}
        \ln a_i b_i^n & if & i = 0  \\
        \ln (a_i b_i^n - a_{i-1} b_{i-1}^n) & if & 1 \leq i \leq r_{\max} . 
    \end{array} \right. 
\end{equation*}
When $i = 0$, 
\begin{equation*}
\ln p_n(i) = \ln a_i  b_i^n = \ln a_i + n \ln b_i,
\end{equation*}
which is linear in $n$ and thus concave in $n$. When $1 \leq i \leq r_{\max}$, we prove that $\ln p_{Z_n}(i)$ is concave in $n$ by showing that its second derivative with respect to $n$ is non-positive: 
\begin{equation*}
\ln p_n(i) = \ln \left(a_i b_i^n - a_{i-1} b_{i-1}^n\right),
\end{equation*}
\begin{equation*}
\frac{\partial \ln p_n (i)}{\partial n} = \frac{a_i (\ln b_i) b_i^n - a_{i-1} (\ln b_{i-1}) b_{i-1}^n}{a_i b_i^n - a_{i-1} b_{i-1}^n},
\end{equation*}
\begin{equation*}
\frac{\partial^2 \ln p_n (i)}{\partial n^2} = -\frac{a_{i-1}a_i b_{i-1}^n b_i^n \left(\ln b_{i-1} - \ln b_i\right)^2}{(a_i b_i^n - a_{i-1} b_{i-1}^n)^2} < 0.
\end{equation*}

\begin{figure}[p]
\centering
\begin{subfigure}{0.48\textwidth}
  \centering
  \includegraphics[scale=0.4]{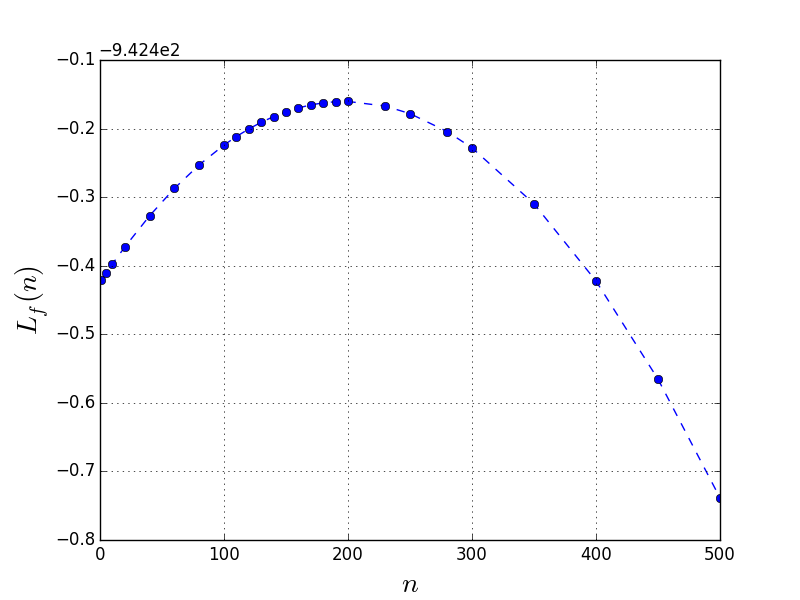}
  \caption{$n_f = 150$}
\end{subfigure}%
\begin{subfigure}{0.48\textwidth}
  \centering
  \includegraphics[scale=0.4]{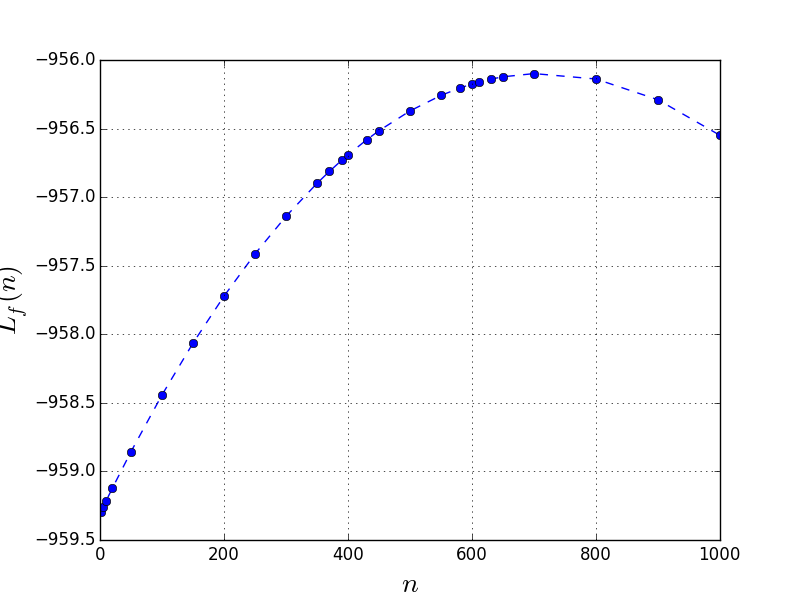}
  \caption{$n_f = 611$}
\end{subfigure}
\begin{subfigure}{0.48\textwidth}
  \centering
  \includegraphics[scale=0.4]{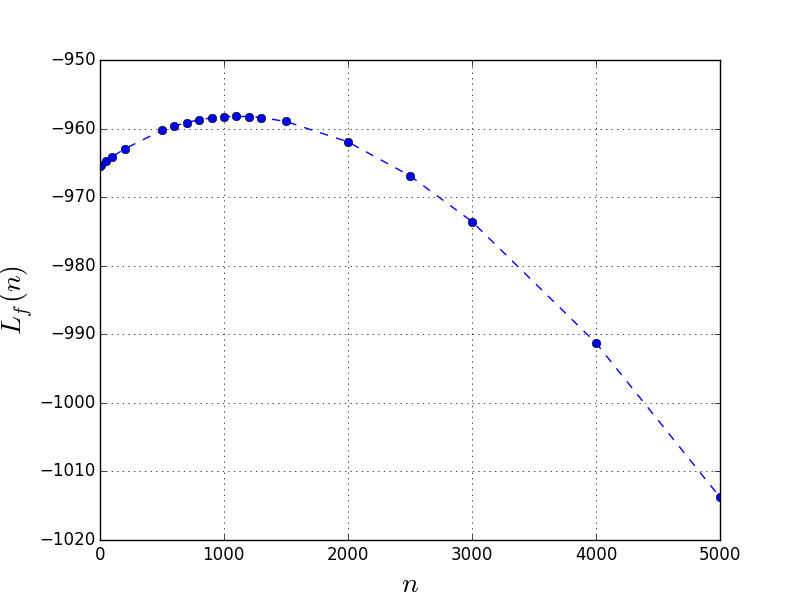}
  \caption{$n_f = 1000$}
\end{subfigure}
\begin{subfigure}{0.48\textwidth}
  \centering
  \includegraphics[scale=0.4]{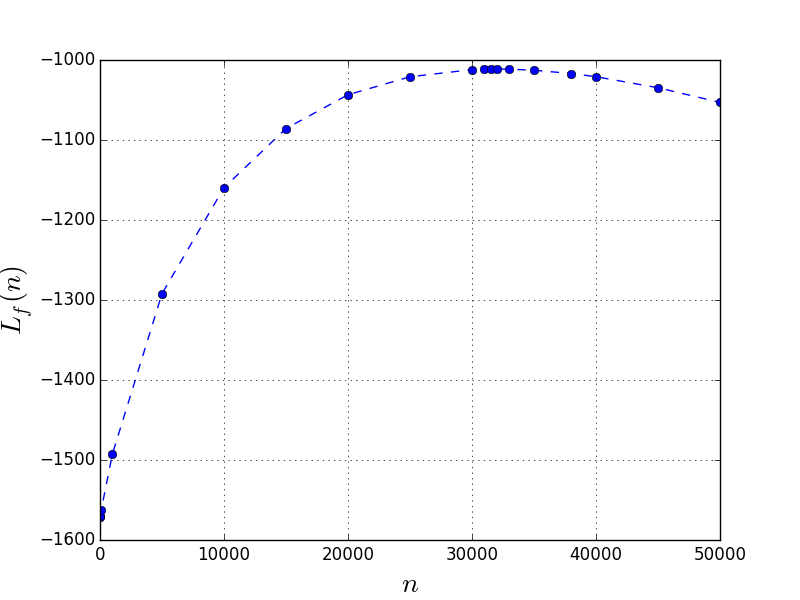}
  \caption{$n_f = 31536$}
\end{subfigure}
\caption{Four sample plots of log likelihood function $L_f(n)$ vs. $n$.}
\label{fig:samples_log_likelihood_vs_n}
\end{figure}
We conclude that $L_f(n)$ is concave in $n$. 

In Figure \ref{fig:samples_log_likelihood_vs_n}, we show the plot of $L_f(n)$ as a function of $n$ for four selected flows whose cardinalities are $150$, $611$, $1000$ and $31536$, respectively. The plots are generated based on the same trace data used for Figure \ref{fig:trace_flow_cardi_dist} and Figure \ref{fig:sample_Z_dist}. The concavity of $L_f(n)$ is obvious. We can also observe that in each plot, the value of $n$ at which the curve peaks is close to the actual flow cardinality $n_f$ (though not exact the same due to randomness and noise). 

\newpage

\subsection{Bi-section search implementation}

We have shown that $L_f(n)$ is concave in $n$. Since $n$ can only take integer values in $\{1, 2, \cdots, n_{\max}\}$,  we can do a bi-section search to find the optimal $n$. 

\begin{framed}
\textbf{Bi-section search implementation of $\widehat{n}_{f,\ML}$}:
\begin{enumerate}
    \item $\lb \leftarrow 1$, $\ub \leftarrow n_{\max}$. 
    \item While $\ub - \lb > 1$:
    \begin{itemize}
        \item $\middleone \leftarrow \floor{\frac{\lb + \ub}{2}}$, $\middletwo \leftarrow \middleone + 1$
        \item if $L_f(\middleone) \geq L_f(\middletwo)$:
            \begin{itemize}
                \item $\lb \leftarrow \lb$, $\ub \leftarrow \middleone$ 
            \end{itemize}
        \item else:
            \begin{itemize}
                \item $\lb \leftarrow \middletwo$, $\ub \leftarrow \ub$
            \end{itemize}
    \end{itemize}
    \item If $L_f(\lb) \geq L_f(\ub)$, return $\lb$; otherwise return $\ub$.
\end{enumerate}
\end{framed}

The pseudo-code above should be self-explanatory. In actual implementation, it is found that the algorithm reaches the optimal $n$ at a faster rate if we replace $\middleone = \floor{\frac{\lb + \ub}{2}}$ with $\middleone = \sqrt{\lb \cdot \ub}$, which is equivalent to $\log \middleone = \frac{1}{2}(\log \lb + \log \ub)$, i.e. a bi-section search on log scale. This faster rate is because, as we already mentioned, most of the flows have small cardinalities. If $n_{\max} = 10^6$, then at the first while loop iteration,  with $\middleone = \floor{\frac{\lb + \ub}{2}}$, we have $\middleone = 5 \times 10^5$; but with $\middleone = \sqrt{\lb \cdot \ub}$ we have $\middleone = 1000$ --- the latter search is quicker if the optimal $n$ is actually small.

With bi-section search, the number of searches required for the estimation of one flow is upper bounded by $\log n_{\max}$. This is not ideal compared to the $\vLL_\theta$ estimator which only needs one search/estimation. However, since the estimation process is off-line, the worst-case time complexity of $\log n_{\max}$ for searching is still acceptable.

\subsection{Summary of the maximum-likelihood estimator}

We wrap up the ideas in the previous sections of this chapter and name this estimator vLL-MLE, meaning that it uses \emph{maximum likelihood estimation} based on \emph{virtual LogLog sketching}. The estimator is summarized as follows. To estimate the cardinality of flow $f$:
\begin{enumerate}
	\item Find the values of the $k$ registers in  $R_f$: $R_f[0], \cdots, R_f[k-1]$.
	\item From the register values in $R$, find the CDF of $Z$ by (\ref{eq:Z_cdf}).
	\item Do a bi-section search on integer value $n$ in the range $\{1, 2, \dots, n_{\max}\}$ to maximize the log likelihood function $L_f(n)$, output this value of $n$ as the estimate.
	\begin{itemize}
		\item For a given $n$, $L_f(n)$ can be evaluated by (\ref{eq:Zn_pmf}), (\ref{eq:Wn_cdf}) and (\ref{eq:log_likelihood_function2}).
	\end{itemize}
\end{enumerate}

\section{Aside: MLE for single-flow estimation} \label{sec:mle_single_flow}

As an aside, we can also use the vLL-MLE estimator for single-flow estimation, with some minor changes. Recall that in single-flow estimation, the goal is to estimate the flow's cardinality $n$ from its $k$ register values. This is similar to the per-flow estimation case, where the goal is to estimate a given flow $f$'s cardinality $n_f$ from its $k$ register values $R_f[0], \cdots, R_f[k-1]$, except that in the single-flow case, there is no background noise in the register values. That is, the vLL-MLE estimator for per-flow estimation can be used for single-flow estimation by simply letting $Z = 0$.

In this case, we do not need the assumptions related to $Z$ anymore. But the Poissonization approximation of $n$ (i.e. assuming $n$ is large) is still necessary. 

To evaluate the performance of the MLE on single-flow estimation, we run experiments to obtain empirical values of its relative standard error for selected values of $k$ and compare it with the HLL estimator. The experiments are performed with $n = 10^7$. The results are summarized in Table \ref{tb:compare_mle_hll_single_flow}.

\begin{table}[!h]
	\centering
	\caption{Compare the accuracy (relative standard error) of the MLE and HLL estimators with selected values of $k$.}
	\begin{tabular}{|c|c|c|c|}
		\hline
		Estimator  & $k=512$ & $k=1024$ & $k=2048$  \\
		\hline
		MLE  &    $0.04609$  & $0.03127$  &  $0.02220$  \\
		\hline
		HLL  &    $0.04585$  & $0.03127$  &  $0.02216$   \\
		\hline
	\end{tabular}
	\label{tb:compare_mle_hll_single_flow}
\end{table}

The experiment results show that the MLE's performance is very close to that of HLL, but not any better. It reinforces the claim that the HLL estimator is near-optimal for single-flow estimation.

\section{Experimental performance evaluation}

In this section we evaluate the performance of the vLL-MLE estimator and compare it with that of the $\vLL_\theta$ estimator introduced in Chapter \ref{ch:generalize_theta}. Since we have shown that $\vLL_{-1}$ (i.e. vHLL) is the best within the family of the $\vLL_\theta$ estimators, we compare vLL-MLE with $\vLL_{-1}$ in particular. Results for each estimator are generated by running experiments on the same $100$ simulated trace files. 

\begin{figure}[!h]
	\centering
	\begin{subfigure}{0.48\textwidth}
		\centering
		\includegraphics[scale=0.4]{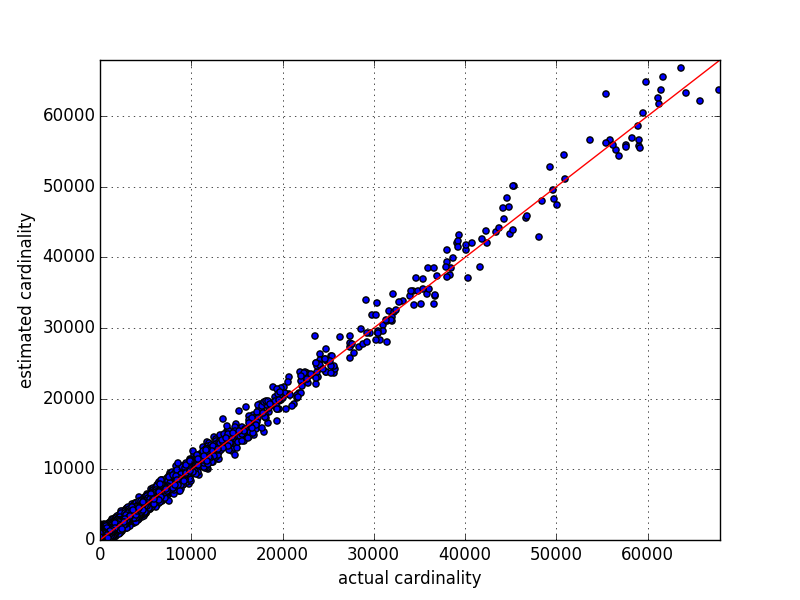}
		\caption{vLL-MLE}
	\end{subfigure}%
	\begin{subfigure}{0.48\textwidth}
		\centering
		\includegraphics[scale=0.4]{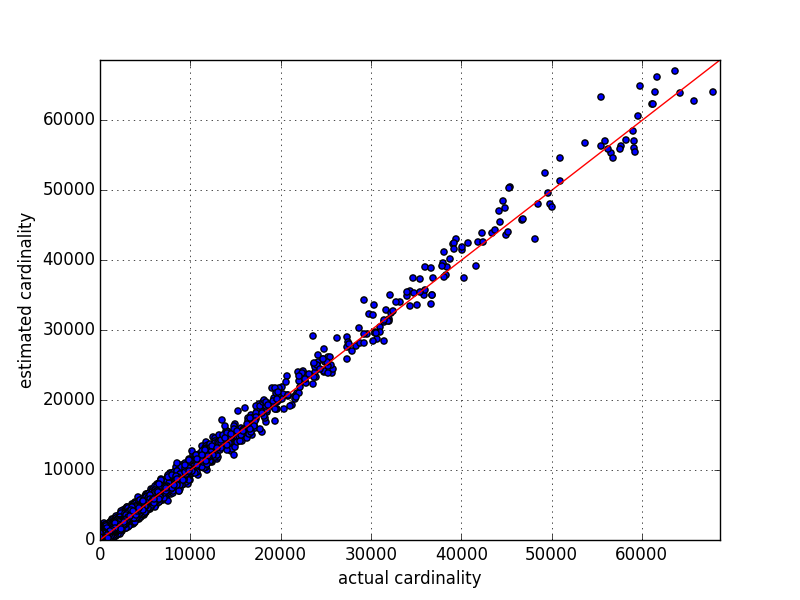}
		\caption{$\vLL_{-1}$}
	\end{subfigure}
	\caption{Plot of estimated cardinalities vs. actual cardinalities for the vLL-MLE and $\vLL_{-1}$ estimators.}
	\label{fig:compare_mle_vhll_scatter_plot}
\end{figure}

Figure \ref{fig:compare_mle_vhll_scatter_plot} shows the estimation results of both estimators by directly plotting the estimated cardinalities vs. the corresponding actual cardinalities. Each point in the graphs represents one flow, with its x-coordinate value being the actual cardinality of the flow and its y-coordinate value being the estimated cardinality. The more clustered the points are to the equality line $y=x$, the more accurate the estimator is. We can see that the two estimators have comparable performances. 

\begin{figure}[!h]
	\centering
	\includegraphics[scale=0.5]{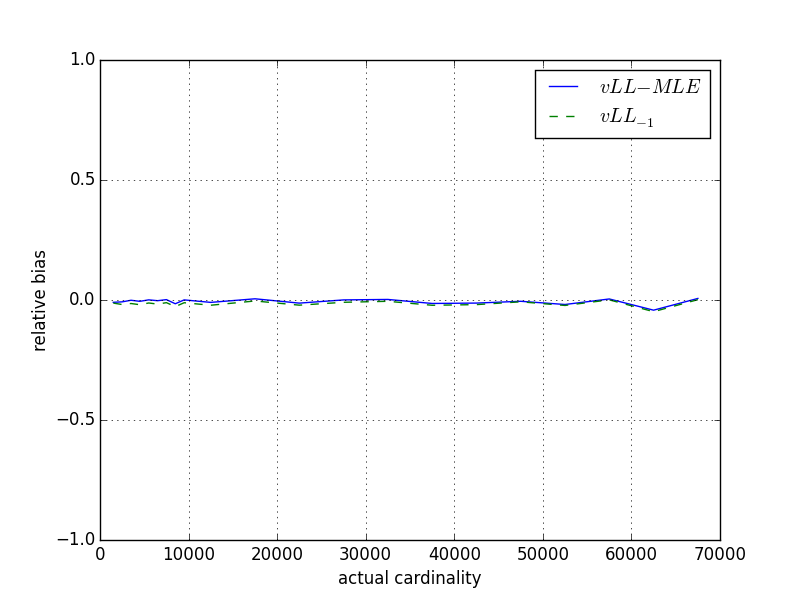}
	\caption{Simulated results on the relative bias of the vLL-MLE and $\vLL_{-1}$ estimators.}
	\label{fig:mle_vs_vhll_on_bias}
\end{figure}

\begin{figure}[!h]
	\centering
	\includegraphics[scale=0.5]{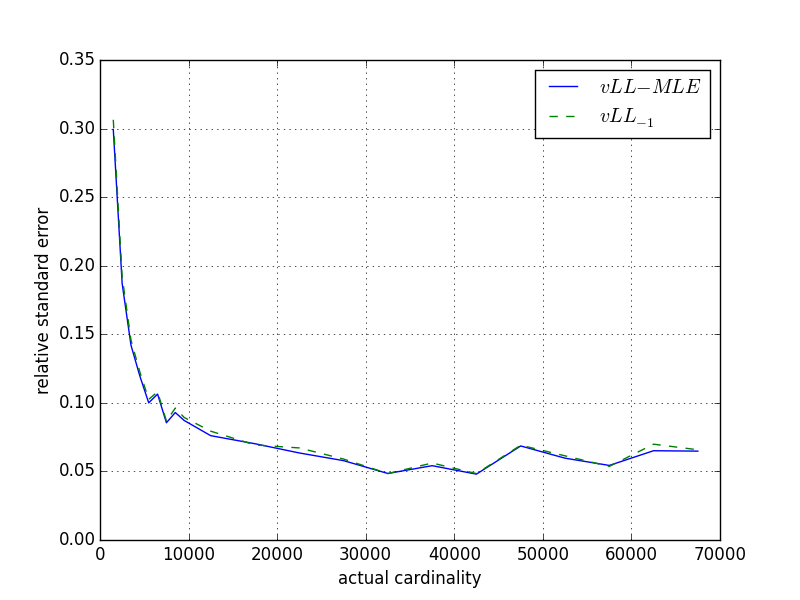}
	\caption{Simulated results on the relative standard error of the vLL-MLE and $\vLL_{-1}$ estimators.}
	\label{fig:mle_vs_vhll_on_std_err}
\end{figure}

In Figure \ref{fig:mle_vs_vhll_on_bias} and Figure \ref{fig:mle_vs_vhll_on_std_err}, we respectively plot the relative bias (defined as $\E\left(\frac{\widehat{n}_f}{n_f}\right) - 1$) and relative standard error (defined as $\frac{\sqrt{\Var(\widehat{n}_f)}}{n_f}$) vs. the actual flow cardinality for both estimators. Since there are not many flows for some cardinalities (especial the large cardinalities), we divide the horizontal axis into bins of width $1000$ for cardinalities $\leq 10000$ and of width $5000$ for cardinalities $ > 10000$. In each bin, we calculate the empirical relative bias and relative standard error of the data and interpolate the values for each bin on the graph to form the plots. Again we see that the two estimators have very similar performances.

In both figures, the curves are plotted for cardinalities larger than $1000$ only. This is for better graph layouts: we found from the experiment results that estimates for flows with cardinalities less than $1000$ are very inaccurate. Accurate estimation for small flows is difficult because of the noise caused by large flows in their register values. 

Figure \ref{fig:mle_vs_vhll_on_bias} shows that both estimators are approximately unbiased for large flows (with cardinalities $> 1000$).  

Figure \ref{fig:mle_vs_vhll_on_std_err} shows that in general, for both estimators, the estimation accuracy improves as the cardinality gets larger. We do observe a slight curving up at the high end of the graph though, meaning that the accuracy stops improving once the actual flow cardinality reaches a certain level (probably $> 40000$).

\begin{figure}[!h]
	\centering
	\includegraphics[scale=0.5]{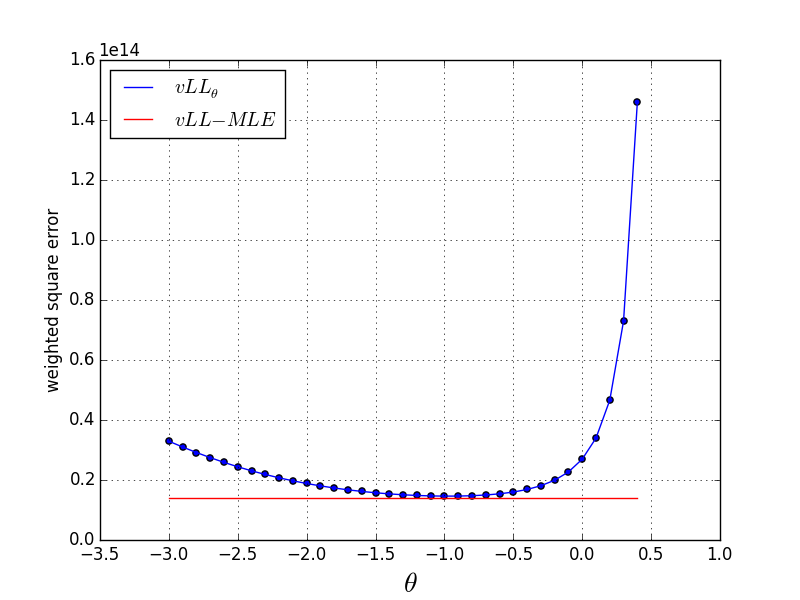}
	\caption{Compare the weighted square error of the vLL-MLE estimator and the $\vLL_\theta$ estimators.}
	\label{fig:wse_vs_theta_plus_mle}
\end{figure}

Finally, in Figure \ref{fig:wse_vs_theta_plus_mle} we compare the weighted square error of the vLL-MLE estimator with that of the $\vLL_\theta$ estimators. Experiment results show that the vLL-MLE estimator outperforms the $\vLL_\theta$ estimator for all values of $\theta$; compared to the $\vLL_{-1}$ estimator, the vLL-MLE estimator has a slight improvement of about $3.5\%$. We conclude that the two estimators, vLL-MLE and $\vLL_{-1}$ (i.e. vHLL), have comparable performances.

\chapter{Conclusions and Future Work} \label{ch:conclusion}

In this thesis we explored two new perspectives on the estimation process of the virtual LogLog algorithm \cite{Xiaoetc2015} for per-flow cardinality estimation:
\begin{itemize}
	\item We showed how the existing vHLL estimator of \cite{Xiaoetc2015} for per-flow estimation can be generalized by introducing a parameter $\theta$, in a similar way in which the HLL estimator for single-flow estimation can be generalized by $\theta$. 
	\item We proposed the vLL-MLE estimator, an alternative approach to the per-flow estimation problem. 
\end{itemize} 

In both cases we provided empirical evidence to show the near-optimality of the vHLL estimator for per-flow estimation. This result is analogous to the near-optimality of the HLL estimator for single-flow estimation \cite{Flajoletetc2007}. 

Results of this thesis are mostly based on simulated experiments. One possible future work is the analysis of theoretical bounds for per-flow cardinality estimation. For example, for a given amount of memory, flow cardinality distribution and number of flows, what is the best possible level of accuracy that can be achieved for per-flow estimation? There has been much research on this direction for single-flow estimation. For example, \cite{ChassaingGerin2011} gives an asymptotic lower bound on the relative standard error of single-flow estimators based on order statistics. It would be useful to obtain similar results for per-flow estimation.  

Another possible direction of future work is on efficient algorithms that identify heavy-hitters (i.e. large flows). With the per-flow estimators discussed in this thesis, one is able to estimate the cardinality of a flow \emph{given} the flow's ID. However, in many applications, the goal is to \emph{identify} large flows, in which case we are not given the IDs of these flows beforehand. One possible approach to this problem is to store all the distinct flow IDs using a separate block of memory and then check each flow one by one; the work in \cite{Yoonetal2009} discusses how all the distinct flow IDs can be stored in main memory. However, with this approach some memory is wasted on flows that are actually small (which do not matter at all). Algorithms that identify large flows and estimate their cardinalities directly and more efficiently are useful in this context.

%

\backmatter

%
\bibliographystyle{IEEE_ECE}
\bibliography{thesisrefs}

\interlinepenalty10000
\begin{thebibliography}{10}
\providecommand{\url}[1]{#1}
\csname url@samestyle\endcsname
\providecommand{\newblock}{\relax}
\providecommand{\bibinfo}[2]{#2}
\providecommand{\BIBentrySTDinterwordspacing}{\spaceskip=0pt\relax}
\providecommand{\BIBentryALTinterwordstretchfactor}{4}
\providecommand{\BIBentryALTinterwordspacing}{\spaceskip=\fontdimen2\font plus
\BIBentryALTinterwordstretchfactor\fontdimen3\font minus
  \fontdimen4\font\relax}
\providecommand{\BIBforeignlanguage}[2]{{%
\expandafter\ifx\csname l@#1\endcsname\relax
\typeout{** WARNING: IEEEtran.bst: No hyphenation pattern has been}%
\typeout{** loaded for the language `#1'. Using the pattern for}%
\typeout{** the default language instead.}%
\else
\language=\csname l@#1\endcsname
\fi
#2}}
\providecommand{\BIBdecl}{\relax}
\BIBdecl

\bibitem{Giroire2009}
F.~Giroire, ``Order statistics and estimating cardinalities of massive data
  sets,'' \emph{Discrete Applied Mathematics}, vol. 157, pp. 406--427, Jan
  2009.

\bibitem{ChassaingGerin2011}
P.~Chassaing and L.~Gerin, ``Efficient estimation of the cardinality of large
  data sets,'' 2011, arXiv:math/0701347v3.

\bibitem{Xiaoetc2015}
Q.~Xiao, S.~Chen, M.~Chen, and Y.~Ling, ``Hyper-compact virtual estimators for
  big network data based on register sharing,'' in \emph{Proceedings of the ACM
  SIGMETRICS 2015}, Portland, Oregon, USA, 2015, pp. 417--428.

\bibitem{EstanVargheseFisk2003}
C.~Estan, G.~Varshese, and M.~Fisk, ``Bitmap algorithms for counting active
  flows on high speed link,'' \emph{IEEE/ACM Transactions on Networking},
  vol.~14, pp. 923--937, Oct 2006.

\bibitem{FlajoletMarin1985}
P.~Flajolet and G.~N. Martin, ``Probabilistic counting algorithms for data base
  algorithms,'' \emph{Journal of Computer and System Sciences}, vol.~31, no.~2,
  pp. 182--209, Sep 1985.

\bibitem{HeuleNunkesserHall2013}
S.~Heule, M.~Nunkesser, and A.~Hall, ``Hyperloglog in practice: Algorithmic
  engineering of a state of the art cardinality estimation algorithm,'' in
  \emph{Proceedings of the EDBT 2013 Conference}, Genoa, Italy, Mar 2013.

\bibitem{Flajoletetc2007}
P.~Flajolet, E.~Fusy, O.~Gandouet, and F.~Meunier, ``Hyperloglog: The analysis
  of a near-optimal cardinality estimation algorithm,'' in \emph{Proceedings of
  AOFA ’07}, 2007, pp. 127--146.

\bibitem{DurandFlajolet2003}
M.~Durand and P.~Flajolet, ``Loglog counting of large cardinalities,'' in
  \emph{European Symposium on Algorithms}, 2003, pp. 605--617.

\bibitem{Gibbons2007}
\BIBentryALTinterwordspacing
P.~B. Gibbons, ``Distinct-values estimation over data streams,'' in \emph{Data
  Streams Management: Processing High-Speed Data Streams}, 2007. [Online].
  Available: \url{http://www.pittsburgh.intel-research.net/people/gibbons/}
\BIBentrySTDinterwordspacing

\bibitem{MetwallyAgrawalAbbadi2008}
A.~Metwally, D.~Agrawal, and A.~E. Abbadi, ``Why go logarithmic if we can go
  linear?: Towards effective distinct counting of search traffic,'' in
  \emph{Proceedings of the 11th International Conference on Extending Database
  Technology: Advances in Database Technology}, ser. EDBT '08.\hskip 1em plus
  0.5em minus 0.4em\relax New York, NY, USA: ACM, 2008, pp. 618--629.

\bibitem{CliffordCosma2010}
P.~Clifford and I.~A. Cosma, ``A statistical analysis of probabilistic counting
  algorithms,'' \emph{Scandinavian Journal of Statistics}, vol.~39, no.~1, pp.
  1--14, Jun 2010.

\bibitem{CAIDA}
``The {CAIDA} {UCSD} anonymized internet traces 2013 on {J}an 17.''
  \url{http://www.caida.org/data/passive/passive_2013_dataset.xml}.

\bibitem{Moetal2014}
Z.~Mo, Y.~Qiao, S.~Chen, and T.~Li, ``Highly compact virtual maximum likelihood
  sketches for counting big network data,'' in \emph{52 Annual Allerton
  Conference}, Allerton House, UIUC, Illinois, USA, Oct 2014.

\bibitem{MitzenmacherUpfal}
M.~Mitzenmacher and E.~Upfal, \emph{Probability and Computing: Randomized
  Algorithms and Probabilistic Analysis}, 1st~ed.\hskip 1em plus 0.5em minus
  0.4em\relax Cambridge, UK: Cambridge University Press, 2005.

\bibitem{AdamicHuberman2002}
L.~A. Adamic and B.~A. Huberman, \emph{Glottometrics}, pp. 143--150, 2002.

\bibitem{Yoonetal2009}
M.~Yoon, T.~Li, S.~Chen, and J.-K. Peir, ``Fit a compact spread estimator in
  small high-speed memory,'' \emph{IEEE/ACM Trans. Netw.}, vol.~19, no.~5, pp.
  1253--1264, Oct 2011.

\end{thebibliography}


\end{document}